\documentclass[
    reprint,
    floatfix,
    superscriptaddress,
    amsmath,amssymb,
    aps,
    pra,
    longbibliography  
]{revtex4-2}

\usepackage[utf8]{inputenc}
\usepackage{graphicx,import}
\graphicspath{{Figures/}}
\usepackage{bm}
\usepackage[colorlinks]{hyperref} 
\usepackage{nameref}
\usepackage[all]{hypcap}
\usepackage[capitalise,nameinlink]{cleveref} 

\usepackage{xcolor}
\usepackage{braket}
\usepackage{siunitx}
\DeclareSIUnit{\rpm}{rpm} 
\DeclareSIUnit\sccm{sccm}
\usepackage{float}
\usepackage{lipsum}

\usepackage{dcolumn}
\usepackage{booktabs}
\usepackage[version=4]{mhchem}

\usepackage{multirow}  

\usepackage{svg} 

\DeclareSIUnit{\bar}{bar}

\definecolor{bluegray}{RGB}{40,180,160}
\definecolor{navygray}{RGB}{110,140,170}
\definecolor{meadowgreen}{RGB}{0,128,0}
\definecolor{coolbrown}{RGB} {165,42,42}

\DeclareSIUnit{\sq}{\Box}

\hypersetup{
    citecolor={navygray}, 
    linkcolor={navygray},
    urlcolor={blue},
}

\newcommand{\revisex}[1]{{}}

\makeatletter
\newcommand*{\balancecolsandclearpage}{%
  \close@column@grid
  \cleardoublepage
  \twocolumngrid
}
\makeatother

\begin{document}
\title{High~Impedance~Granular~Aluminum~Ring~Resonators}

\author{Mahya~Khorramshahi}
\affiliation{IQMT,~Karlsruhe~Institute~of~Technology,~76131~Karlsruhe,~Germany}

\author{Martin~Spiecker}
\affiliation{IQMT,~Karlsruhe~Institute~of~Technology,~76131~Karlsruhe,~Germany}
\affiliation{PHI,~Karlsruhe~Institute~of~Technology,~76131~Karlsruhe,~Germany}

\author{Patrick~Paluch}
\affiliation{IQMT,~Karlsruhe~Institute~of~Technology,~76131~Karlsruhe,~Germany}
\affiliation{PHI,~Karlsruhe~Institute~of~Technology,~76131~Karlsruhe,~Germany}

\author{Simon~Geisert}
\affiliation{IQMT,~Karlsruhe~Institute~of~Technology,~76131~Karlsruhe,~Germany}

\author{Nicolas~Gosling}
\affiliation{IQMT,~Karlsruhe~Institute~of~Technology,~76131~Karlsruhe,~Germany}

\author{Nicolas~Zapata}
\affiliation{IQMT,~Karlsruhe~Institute~of~Technology,~76131~Karlsruhe,~Germany}

\author{Lucas~Brauch}
\affiliation{IQMT,~Karlsruhe~Institute~of~Technology,~76131~Karlsruhe,~Germany}

\author{Christian~K\"ubel}
\affiliation{INT,~Karlsruhe~Institute~of~Technology,~76131~Karlsruhe,~Germany}
\affiliation{KNMFi, Karlsruhe Institute of Technology, 76131 Karlsruhe, Germany}
\affiliation{Technische Universit\"at Darmstadt, 64289 Darmstadt, Germany}

\author{Simone~Dehm}
\affiliation{INT,~Karlsruhe~Institute~of~Technology,~76131~Karlsruhe,~Germany}

\author{Ralph~Krupke}
\affiliation{IQMT,~Karlsruhe~Institute~of~Technology,~76131~Karlsruhe,~Germany}
\affiliation{INT,~Karlsruhe~Institute~of~Technology,~76131~Karlsruhe,~Germany}

\author{Wolfgang~Wernsdorfer}
\affiliation{IQMT,~Karlsruhe~Institute~of~Technology,~76131~Karlsruhe,~Germany}
\affiliation{PHI,~Karlsruhe~Institute~of~Technology,~76131~Karlsruhe,~Germany}

\author{Ioan~M.~Pop}
\affiliation{IQMT,~Karlsruhe~Institute~of~Technology,~76131~Karlsruhe,~Germany}
\affiliation{PHI,~Karlsruhe~Institute~of~Technology,~76131~Karlsruhe,~Germany}
\affiliation{Physics~Institute~1,~Stuttgart~University,~70569~Stuttgart,~Germany}

\author{Thomas~Reisinger}
\email{thomas.reisinger@kit.edu}
\affiliation{IQMT,~Karlsruhe~Institute~of~Technology,~76131~Karlsruhe,~Germany}

\date{\today}

\begin{abstract}
Superconducting inductors with impedance surpassing the resistance quantum, i.e., superinductors, are important for quantum technologies because they enable the development of protected qubits, enhance coupling to systems with small electric dipole moments, and facilitate the study of phase-slip physics. We demonstrate superinductors with densely packed meandered traces of granular aluminum (grAl) with inductances up to \SI{4}{\micro \henry}, achieving impedances exceeding \SI{100}{\kilo \ohm} in the \SIrange{4}{8}{\giga \hertz} range. Ring resonators made with grAl meandered superinductors exhibit quality factors on the order of \(10^5\) in the single-photon regime, and low non-linearity on the order of tens of Hz. Depending on the grAl resistivity, at \SI{10}{\hertz}, we measure frequency noise spectral densities in the range of \(10^2\) to \(10^3\) \SI{}{\hertz/\sqrt{\hertz}}. In some devices, in the single-photon regime, we observe a positive Kerr coefficient of unknown origin. Using more complex fabrication, the devices could be released from the substrate, either freestanding or suspended on a membrane, thereby further improving their impedance by a factor of three.
\end{abstract}



\maketitle
\clearpage

In circuits with high characteristic impedance \( Z_\mathrm{C} \), zero-point voltage fluctuations are enhanced~\cite{devoretCircuitQED2007, burkardSpinQubits2023}, \(V_{\text{ZPF}} \propto \sqrt{Z_\mathrm{C}}\), enabling coupling of superconducting microwave devices to other systems, e.g. phonons via direct piezoelectric coupling~\cite{arrangoiz-arriolaCoupling2018} or parametric electromechanical interactions~\cite{bozkurtQuantum2023}, molecular qubits~\cite{andreCoherent2006}, as well as small electric dipoles and spins in quantum dots~\cite{burkardSpinQubits2023}. At the high end of \( Z_\mathrm{C} \), inductors with zero DC resistance and impedance exceeding the resistance quantum, \(Z_C > R_Q = h/(2e)^2 \approx \SI{6.45}{\kilo\ohm}\), are known as superinductors~\cite{manucharyanSuperinductor2012, maslukMicrowave2012}. They enable the realization of protected qubit architectures, such as fluxonium \cite{manucharyanFluxonium2009, popCoherent2014,grunhauptGranular2019, pita-vidalFluxonium2020, nguyenFluxonium2022, peruzzoGeoQubits2021, kalashnikovBifluxon2020} and  \(0-\pi\) qubits~\cite{brooks0-pi2013, groszkowski0pi02018, gyenis0pi2021}, by delocalizing the wavefunction and mitigating flux noise. The high inductive reactance of superinductors is also essential for studying flux-tunneling-induced phase slips, facilitating the observation of dual Shapiro steps~\cite{shaikhaidarovCurrentSteps2022, cresciniDual2023,  kaapShapiro2024}.

Long Josephson junction arrays are a widely used option for achieving high-impedance with low-loss~\cite{manucharyanSuperinductor2012, maslukMicrowave2012}, and their impedance can exceed hundreds of \(\si{\kilo\ohm}\) when suspended in vacuum~\cite{pechenezhskiySuperconding2020, jungerSuspended2025}. However, these arrays—particularly in suspended configurations—are complex to fabricate and introduce unwanted non-linearity~\cite{frascaNbN2023}. Alternatively, geometric inductors optimized into suspended spiral shapes offer high linearity but remain challenging to fabricate~\cite{peruzzoSurpassing2020, medahinneMagnetic2024}. 
\enlargethispage{\baselineskip}

An ideal superinductor technology would combine low dissipation, straightforward fabrication, and compatibility across diverse operational environments, including high magnetic fields, as required for example in spin~\cite{burkardSpinQubits2023}, Andreev~\cite{pita-vidalAndreev2024}, or nanowire qubit devices~\cite{larsenNanowireBQubit2015, valentiniNanowireQubit2021, luthiNanowireTransmon2018}. 

In this context, disordered superconductors are particularly attractive because they can be densely patterned using standard lithographic techniques, while contributing substantial kinetic inductance dominating the total inductive response. Among the various candidate materials, several have been employed to fabricate devices exhibiting characteristic impedance \( Z_\mathrm{C} \) higher than \SI{1}{\kilo\ohm} and internal quality factors \(Q_i\) above \(10^5\) in the single-photon regime. These include NbN~\cite{niepceHigh2019, yangNbTiNResonator2024, frascaNbN2023}, TiN~\cite{shearrowAtomic2018, aminLossMechanisms2022, vissersLowLossTiN2010}, NbTiN~\cite{samkharadzeHigh2016, krollNbTiN2019, mullerNbTiNquality2022}, NbAlN~\cite{gaoNbAlN2022}, and grAl~\cite{rotzingergral2017, grunhauptGranular2019, kamenovMeandered2020, janikStrong2024, guptaLowLoss2024}. The associated self-Kerr non-linearity can be tuned from approximately \(10~\si{\hertz}\) to \(10^5~\si{\hertz}\) for both NbN~\cite{frascaNbN2023, xukerrNbN2023} and TiN~\cite{joshiKerrNonlinearity2022}, and from \(10^{-2}~\si{\hertz}\) to \(10^6~\si{\hertz}\) for granular aluminum (grAl)~\cite{maleevaCircuit2018, winkelKerrGral2020}. In a recent study~\cite{Roy2025Mar}, the authors compare NbN and grAl devices, highlighting the following trade-off: NbN offers superior magnetic-field resilience, beneficial for hybrid circuit quantum electrodynamics, while grAl is better suited to low-field regimes requiring high impedance and strong non-linearity.

Here, we utilize grAl to explore the high-impedance frontier with meandered trace ring resonators. The meandering design is practical for achieving high-impedance as it allows for long inductive lines within a small area, maximizing inductance while minimizing capacitance~\cite{kamenovMeandered2020}. We demonstrate ring resonators with impedance above \SI{100}{\kilo \ohm} in the technologically relevant frequency range of \SIrange{4}{8}{\giga \hertz}. These resonators achieve quality factors around \(10^5\) in the single-photon regime, with non-linearity in the tens of \(\si{\hertz}\). Frequency noise spectral densities at \SI{10}{\hertz} range from \(10^2\) to \(10^3~\si{\hertz}/\sqrt{\si{\hertz}}\), depending on grAl resistivity.

\begin{figure}[!t]
\includegraphics[width=\columnwidth]{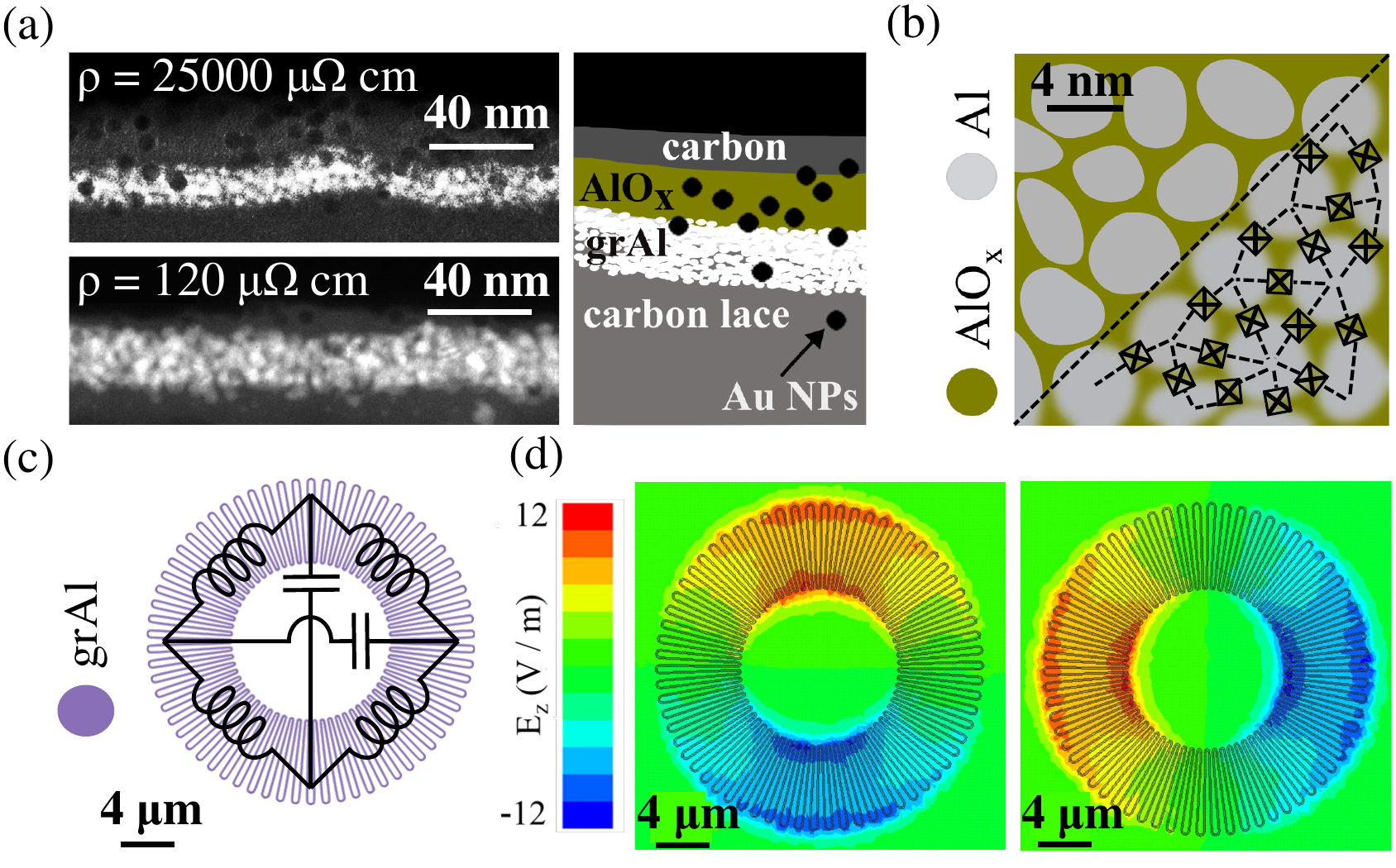}
\caption{\textbf{Granular Aluminum (grAl) Ring Resonators Design.}
(a) Energy-filtered transmission electron microscopy images of high resistivity grAl (top, \(\approx\SI{25000}{\micro\ohm\centi\meter})\) and low resistivity grAl (bottom, \(\approx\SI{120}{\micro\ohm\centi\meter}\)).  The energy filter at \(\SI{15}{\electronvolt}\)  highlights the aluminum volume plasmon. The viewing angle is tilted by \SI{72}{\degree} to the deposition direction for both images. 
The right-hand schematic is there to aid interpretation of the images. The grAl was deposited on carbon laces visible at the bottom of the films. The dark particles are gold nanoparticles (Au NPs) introduced as fiducial markers for tracking the sample during rotation in tomography experiments.
(b) Schematic of granular aluminum, illustrating aluminum grains (gray) encircled by aluminum oxide barriers. This microscopic configuration forms an effective array of Josephson junctions, which gives high kinetic inductance and tunable non-linearity.
(c) Design of the ring resonator and its equivalent circuit. The densely packed meander trace maximizes the ratio of inductance to self-capacitance. The resonator's symmetric design results in two degenerate modes, each involving one effective capacitor shunted by two parallel inductors.
(d) Rendering of the electric field distribution to illustrate the charge distribution within the resonator. The modes are similar to dual modes in Aluminum ring resonators in Ref.~\cite{minevPlanar2013}.}
\label{fig_design}
\end{figure}

GrAl is a nanocomposite of \ce{Al} grains in an aluminum oxide matrix~(see Fig.~\ref{fig_design}(a)), forming a self-assembled 3D network of Josephson junctions (see Fig.~\ref{fig_design}(b)). It is superconducting for a wide range of normal state resistivities \(\rho\) ranging from 1 to \SI{10000}{\micro\ohm\centi\meter}~\cite{levy-bertrandElectrodynamicgral2019}, which can be adjusted via the oxygen partial pressure during deposition. This tunability enables precise control over its kinetic inductance \(L_\mathrm{k}\) allowing the realization of inductors predominantly governed by kinetic inductance, with a kinetic inductance fraction \(\alpha = L_\mathrm{k}/(L_\mathrm{k} + L_\mathrm{g})\) approaching unity, where \(L_\mathrm{g}\) is the geometric inductance. 

The kinetic inductance \( L_\mathrm{k} \) of a superconducting trace of width \( w \) and length \(\ell\) is given by~\cite{tinkhamintroduction2004,annunziataSuperconducting2010}:
\begin{equation}    
L_\mathrm{k} = L_\square  \frac{ \ell}{w} = ~\frac{ \hbar R_\mathrm{n}}{\mathcal{C} k_\mathrm{B} T_\mathrm{c} \pi}  \frac{ \ell}{w},
\label{eq:kinetic_inductance}
\end{equation}

 where \(L_\square\) is the kinetic inductance per square, \( k_\mathrm{B} \) is the Boltzmann constant, \( T_\mathrm{c} \) is the superconducting critical temperature, and \( R_\mathrm{n} \) is the normal-state sheet resistance. The superconducting gap is linked to the critical temperature by \(\Delta(0) = \mathcal{C}~k_\mathrm{B} T_\mathrm{c}\) with \(\mathcal{C} = 1.76\) for grAl resistivity below \SI{2}{\milli\ohm\centi\meter} and \(\mathcal{C} \approx 2.1\) for the high resistivity range~\cite{prachtCooperPairing2016, levy-bertrandElectrodynamicgral2019}.  

In order to increase the characteristic impedance of the resonator $ Z_\mathrm{C}=\sqrt{L/C}$, where $L=L_\mathrm{k} + L_\mathrm{g}$ and $C$ is the resonator's total capacitance, it is essential to optimize the resonator design to minimize $C$ and maximize $L$.
This is achieved by employing a meandered ring resonator (see Fig.~\ref{fig_design}(c)), where the capacitance is reduced by decreasing the ring’s outer radius \(r_\text{out}\) and using the smallest possible meander pitch \( p \).
Since $\alpha$ is close to unity, $L$ is predominantly enhanced by increasing the kinetic inductance $L_\mathrm{k}$ through two mechanisms: increasing the number of squares in the trace, by increasing the length $\ell$ and reducing the width $w$ of the resonator, and raising the sheet resistance $R_\mathrm{n}$, which can be achieved by using higher-resistivity films and by reducing film thickness.

As shown in Fig.~\ref{fig_design}(d), when the ring is closed, the symmetric design of this distributed resonator results in two fundamental modes. The effective circuit model for each mode comprises a capacitor shunted by two parallel inductors \(L/4\). 

\begin{figure*}[t!]
\includegraphics[width=6.8in]{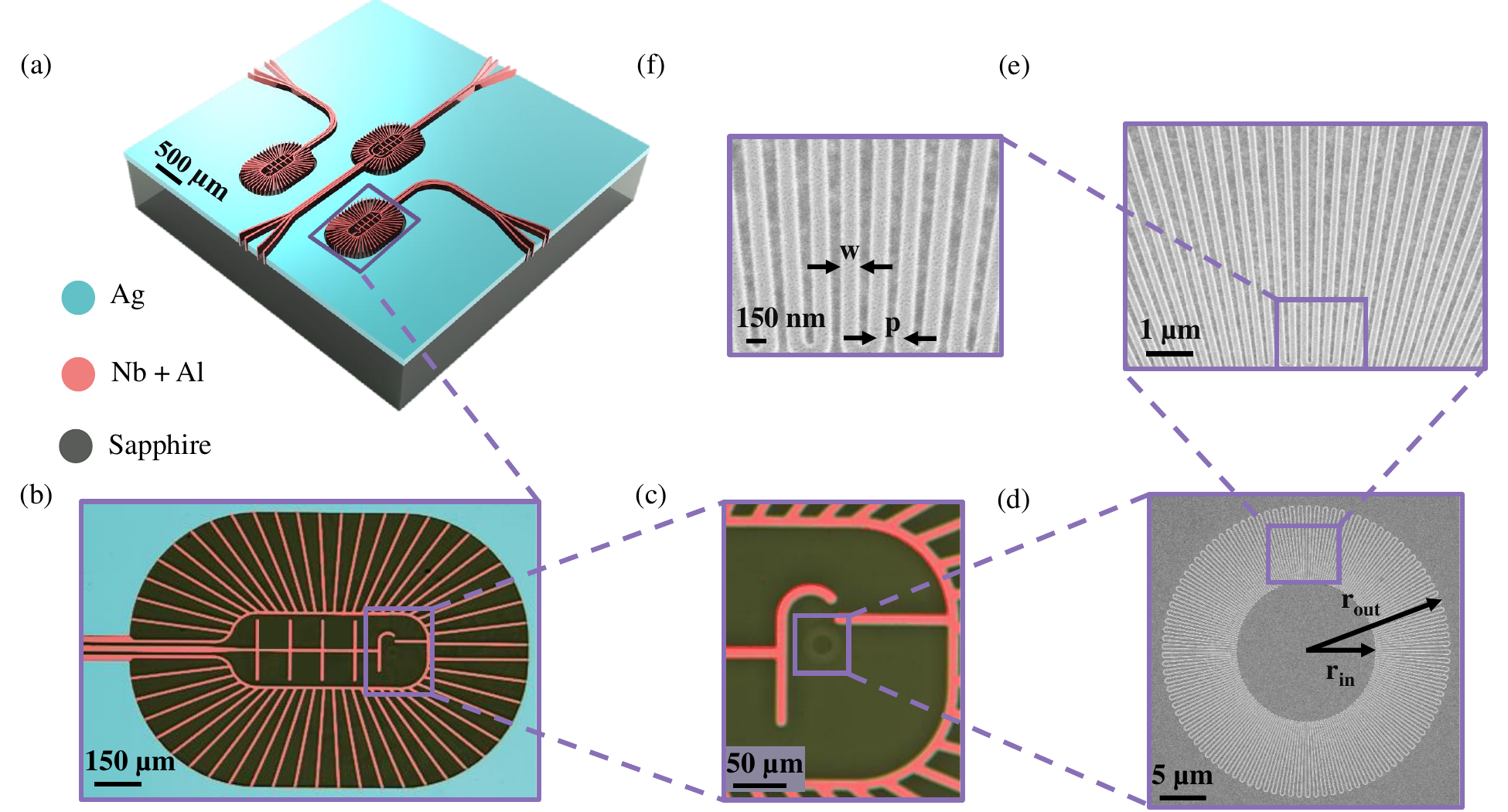}
\caption{\textbf{Ring Resonators in CPW Architecture.} (a) Three-dimensional rendering of the chip layout on a sapphire substrate featuring multiple CPWs, enabling the measurement of various devices, and a surrounding ground plane to suppress parasitic modes and ensure a uniform magnetic field. The ground plane mostly consists of silver, which avoids flux trapping and damps microwave slot modes and phonons~\cite{henriquesPhonon2019}. The central line of the CPW and the regions close to it consist of thin films of \SI{15}{nm} niobium capped by \SI{5}{nm} of aluminum. The \(5\,\mathrm{nm}\) Al capping layer is used to enable clean galvanic contact with the ground in overlapping regions, leveraging the argon milling recipe introduced and tested in Ref.~\cite{grunhauptArgonIonBeam2017}.
(b) Optical false colour image of the end section of the CPW coupled to several ring resonators. The central line splits into either rectangular coupling strips or (c) a semicircle at its end. Both the central line and the ground plane in this region consist of \SI{10}{\micro \meter} wide superconducting strips of \ce{Nb + Al} (see panel (a)), to minimize flux trapping.
(d) Scanning Electron Microscope (SEM) images of a granular aluminum ring resonator, highlighting the densely packed meander traces achieved through a single-step e-beam lithography process followed by lift-off. (e, f) SEM images of the meandering lines at increasing magnification.}
\label{fig_design_details}
\end{figure*}

The resonators are capacitively coupled to a common \(\SI{50}{\Omega}\) coplanar waveguide (CPW) for reflection measurements, as shown in Fig.~\ref{fig_design_details}. The entire structure is integrated within a silver ground plane, which suppresses parasitic modes and ensures a uniform magnetic field across the chip. Near the end of each CPW we push the normal ground plane away from the central pin, in order to reduce dissipation. In the proximity of the devices, the ground plane consists of narrow superconducting lines connected to the normal plane, which avoid the creation of superconducting loops and associated flux trapping. The configuration of the end of CPW feed line enables flexible coupling to the small rings. It intentionally creates different coupling strengths for the two modes, thereby breaking the symmetry and lifting the degeneracy of the two fundamental modes. Positioning the ring near the semicircular structure at the end of the feed line (see Fig.~\ref{fig_design_details}(c)) results in strong coupling for both modes, while placing it within a square structure visible in the center of Fig.~\ref{fig_design_details}(b) leads to uneven coupling of the modes to the feed line.  The compact design allows for the integration of three separate CPW sets on a single chip, each isolated by the ground plane, enabling the measurement of multiple ring resonators with varying design parameters. 

The rings are patterned using a single-step electron beam lithography process followed by lift-off on a c-plane sapphire substrate with zero-angle evaporation deposition. See Figs.~\ref{fig_design_details}(d-f) for images of one of the fabricated rings. The pitch values \(p\) range from \SIrange{200}{520}{\nano\meter}, with widths \(w\) between 150 and \SI{170}{\nano\meter}, and film thicknesses \(t\) from 20 to \SI{30}{\nano\meter}. We wire-bond each chip to a copper housing that is surrounded by magnetic and infrared radiation shielding barrels made of copper/aluminum and \ce{mu}-metal (see Appendix~\ref{app:setup}). The assembly is connected to the mixing chamber (MXC) stage of a dilution cryostat, operating at $\approx \SI{20}{mK}$.

\begin{figure}[!t]
\includegraphics[width = 1\columnwidth]{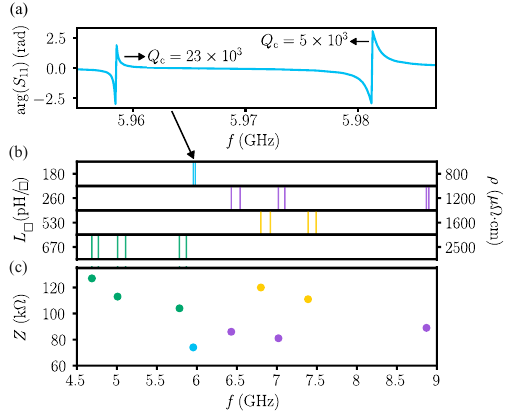}
\caption{\textbf{Frequency and Impedance of Granular Aluminum Superinductors.}
(a) The graph shows the phase response of one of the doublet ring resonances with the frequency separations of approximately \SI{25}{\mega\hertz}, pointing to the asymmetry due to the difference in coupling capacitance of the two modes and fabrication imperfections.
(b) The frequency spectra of the fabricated and measured granular aluminum ring resonators. Each row corresponds to an inductance per square and the corresponding resistivity as determined from equation~\ref{eq:kinetic_inductance}. 
(c) The graph determines the impedance of each ring resonator up to the first fundamental mode. The filled circular markers represent the measured impedance of the fabricated and characterized rings, with the highest impedance recorded at \(\SI{127}{\kilo\Omega}\) at a resonant frequency of \SI{4.6}{\giga\hertz}. These values are obtained using the relation \(Z = 2 \pi f_0 L\). }
\label{fig_impedance_frequency}
\end{figure}

In Fig.~\ref{fig_impedance_frequency}(a), we show a typical phase response of a ring resonator in a microwave reflection measurement. The two fundamental modes are clearly visible, separated by approximately \SI{25}{\mega\hertz}, with their respective coupling quality factors \(Q_c\) differing by approximately a factor of four. 

A summary of the measured fundamental modes frequencies for nine resonators is shown in Fig.~\ref{fig_impedance_frequency}(b), for grAl resistivities between \(\SI{800}{\micro\ohm\centi\meter}\) and \(\SI{2500}{\micro\ohm\centi\meter}\). By performing finite element simulations for rings with varying \(\ell\) and matching the simulated frequencies with measurements, we deduce the inductance per square for each grAl film, ranging from 180 to 670 \SI{}{\pico\henry/\text{sq}} (\(L_{\square}\)).

We determine the impedance up to the first fundamental mode \(f_0\) from the equation \(Z = 2 \pi f_0 L\), where \(L\) denotes the inductance of the entire ring ~\cite{peruzzoSurpassing2020, pechenezhskiySuperconding2020}, as summarized in Fig~\ref{fig_impedance_frequency}(c). 
The highest impedance, \(\SI{127}{\kilo\ohm}\), was achieved for a ring with a sheet inductance of \SI{670}{\pico\henry/\text{sq}}, and specific dimensions of pitch \SI{325}{nm}, \( r_{\text{in}} = \SI{6.7}{\micro\meter} \), \(r_{\text{out}} = 2~r_{\text{in}}\), and thickness \SI{20}{nm}. A detailed summary of all measured devices is provided in  Table~\ref{table3_Design} in Appendix~\ref{app:design_and_extracted_parameters}.

\begin{figure}[t]
\includegraphics[width = 1\columnwidth]{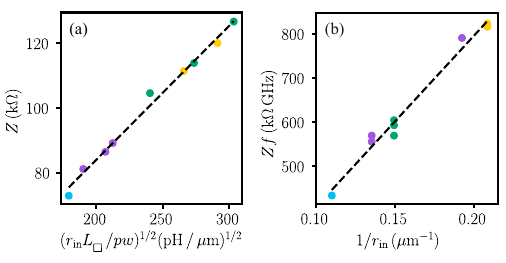}
\caption{\textbf{Measured impedance scaling with geometry and resistivity.}
Plot of measured \(Z\) in panel (a) and \(Zf\) in panel (b) as a function of \((r_{\text{in}}L_{\square}/pw)^{1/2}\) and \(1/r_{\text{in}}\), respectively, confirming their linear dependence (cf. dashed lines), consistent with Eq.~\ref{EqModelZ}. The color code in both panels is consistent with Fig.~\ref{fig_impedance_frequency}. Details for each resonator are given in the supplementary Table~\ref{table3_Design}.}
\label{fig_impedance_model}
\end{figure}

In a simplified model (see Appendix~\ref{app:model}), we find that the impedance and lowest mode frequency scale with the layout parameters and sheet inductance as follows:
\begin{equation}    
Z \propto \sqrt{ \frac{r_{\text{in}} L_{\square}}{p w} } , \, f_0 \propto \sqrt{\frac{p w}{r_{\text{in}}^3 L_{\square}}}.
\label{EqModelZ}
\end{equation}    
Indeed, as shown in Fig.~\ref{fig_impedance_model}(a), the impedance for our measured resonators follows this scaling. Notice that \(Zf_0\propto 1/r_{\text{in}}\), as illustrated in Fig.~\ref{fig_impedance_model}(b), which means that in order to maximize the impedance at a given frequency, we need to minimize \( r_{\text{in}}\) and compensate the resulting change in frequency by decreasing the ratio ${p w}/{L_{\square}}$. The lowest achievable value for the ratio ${p w}/{L_{\square}}$ depends on the grAl material properties and the finesse of the lithography. The maximum value for \(L_{\square}\) is limited by the resistivity threshold for the grAl superconducting-to-insulating transition~\cite{prachtCooperPairing2016, levy-bertrandElectrodynamicgral2019}, in the range of $\SI{10}{\milli\ohm\centi\meter}$, and by the thinnest continuous and stable films. For grAl films below \SI{20}{nm} we observe significant fluctuations in their resistivity between different cooldowns, indicating that structural inhomogeneities and instabilities play a dominant role. We found that devices with \(w \geq \SI{150}{nm}\) always resulted in stable devices, which was not the case for \(w \approx \SI{60}{nm}\), likely due to inhomogeneities in the wires (see Fig.~\ref{fig_design}(a)). The smallest pitch we could achieve with our current electron-beam lithography was \SI{200}{nm} (see Appendix~\ref{app:fabrication}). Taking these considerations into account, we believe that a fine-tuned optimal grAl device operational in the 4-8 \si{\giga\hertz} range can reach impedance values in the range of \SI{200}{\kilo\ohm} (see Appendix~\ref{app:Numerical_Sim}) on Silicon or Sapphire substrates, and this value can be increased by a factor of $\approx 3$, to exceed \SI{0.5}{\mega\ohm} for suspended devices. 

\begin{figure*}[t]
\begin{center}
\includegraphics[width=6.67in]{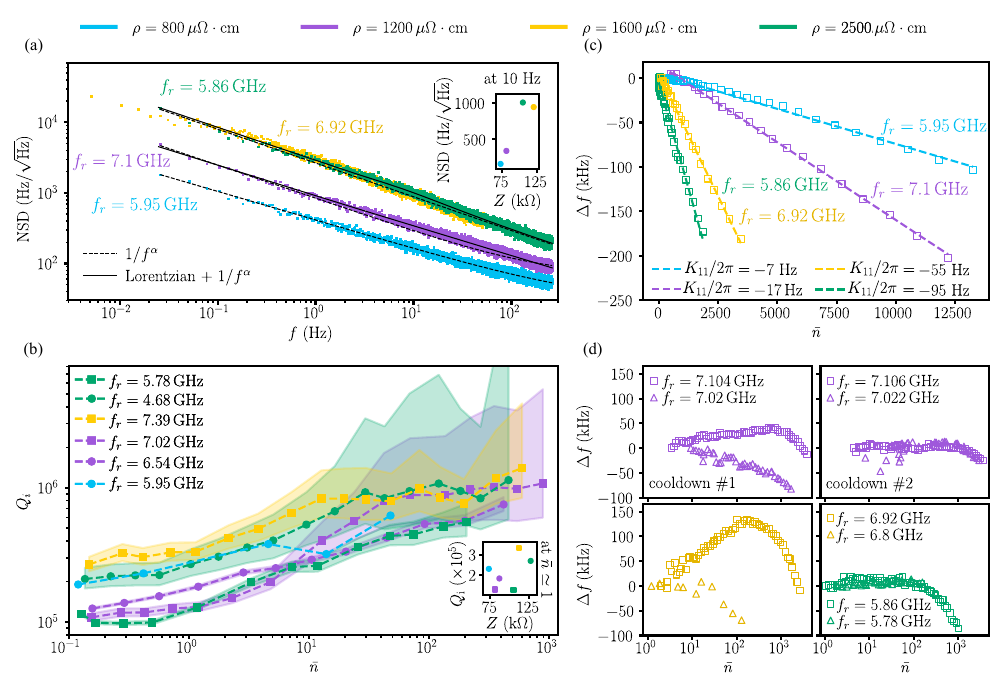}
\caption{\textbf{Resonators Performance.}
(a) Noise spectral density \(S(f)\) as a function of frequency \(f\) for grAl resonators. The spectra were computed from time traces and are shown for four different resistivity values. The data fit the model \(S(f) = S_0 + S_1/f^\alpha\), with \(\alpha\) ranging from 0.8 to 0.97. The two curves corresponding to resistivity values \(\rho = 2500 \, \si{\mu\Omega} \cdot \si{\centi\meter}\) and \(\rho = 1200 \,\si{\mu\Omega} \cdot \si{\centi\meter}\) plotted in green and purple, respectively, exhibit behavior that is better described by a Lorentzian model combined with a \(1/f^{\alpha}\) dependence at higher frequencies. The inset displays the impedance values corresponding to the frequency noise spectral densities measured at \(\SI{10}{\hertz}\). (b) Intrinsic quality factors \( Q_i \) plotted against average photon numbers in the granular aluminum resonators up to the bifurcation threshold \(\bar{n}_{\text{max}}\). The average number of photons is given by \(\bar{n} = 2Q_cP_{\text{cold}}/\omega_0^2\), where  $P_{\text{cold}}$ is the incident microwave power at the device~\cite{grunhauptgralLoss2018}. The shaded intervals indicate the uncertainty range due to Fano interference~\cite{riegerFano2023}. For better visibility, we do not show this uncertainty range for the data set in blue, as it approximately corresponds to an order of magnitude. The inset presents impedance values derived from single-photon \( Q_i \) measurements. (c) The shift in resonance frequency as a function of the average photon number for four different rings with varying resistivity is shown. The self-Kerr coefficient \(K_{11}\) was determined by fitting the frequency shift data to a linear model. Higher resistivity grAl samples exhibit higher \(K_{11}\), consistent with theoretical expectations that \(K_{11}\) is proportional to \(\rho/V\), where \(\rho\) is resistivity and \(V\) is the volume~\cite{maleevaCircuit2018}. (d) Anomalous shifts in resonance frequency with photon number. In some ring resonators, we observe an exponential shift to higher frequency at low photon numbers. In the top two panels, the purple data points show the two modes of the same ring resonator measured during two different cooldowns: In the first, we observe the exponential shift to a higher frequency for the higher frequency mode, while the lower frequency mode almost symmetrically shifts, but to a lower frequency. This occurs before the usual Kerr non-linearity at higher photon numbers documented in (c). In the second cooldown, the exponential shifts are no longer observed. In the bottom left panel, a similar anomalous shift is observed for a resonator with higher resistivity grAl. Next to it, we show an example of a high resistivity grAl resonator that does not show this anomalous shift.}
\label{fig_resonators_performance}
\end{center}
\end{figure*}

We calculate the noise spectral density \( S(f)\) by monitoring the resonant frequency over time and applying the method used in~\cite{bartlettnsd1948}. As shown in Fig.~\ref{fig_resonators_performance}(a), the low-frequency noise exhibits a \(1/f\) dependence. 
We note that \( S(f)\) evaluated at a frequency of \SI{10}{\hertz} is observed in the range of \SIrange{200}{1000}{\hertz\per\sqrt{\hertz}}, which is one to two orders of magnitude larger than for grAl resonators of similar thickness and resistivity, but lower impedance~\cite{valentiKineticInductance2019}. In addition, these values are two orders of magnitude larger compared to aluminum resonators~\cite{burnettNoise2018}.  
The frequency-independent noise component at high-frequency (white noise), modeled by a constant amplitude \( S_0 \), is due to the measurement setup. It is nearly negligible in our data. 
To model the low-frequency noise, we use \( S_{1/f} \propto f^{-\alpha} \), where \( \alpha \approx 1 \).
Fitting this model to the measurements shown in Fig.~\ref{fig_resonators_performance}(a), \( \alpha \) ranges from \( \approx 0.8 \) to \( \approx 0.97 \), which could be due to contributions from other noise sources. 
Note that the three data sets with the highest resistivity exhibit visible deviations from pure \( S(f)\) dependence, which is compatible with an added Lorentzian component. This suggests the presence of random telegraph noise, which decays exponentially over a characteristic timescale given by the position of the peak in the noise spectrum\cite{schlornoise2019}. Note that in our case, the peak is less pronounced than in Ref.~\cite{kristenFluctuations2023}.

To evaluate loss in the resonators, we measure the complex \( S_{11} \) signal and extract the internal quality factor \( Q_i \) and the coupling quality factor \( Q_c \) using the method in Ref.~\cite{riegerFano2023}. To minimize Fano interference systematic errors, we design \( Q_c \) comparable to \( Q_i \). In Fig.~\ref{fig_resonators_performance}(b) we plot the fitted \( Q_i \) values as a function of photon number \(\bar{n}\). In the single-photon regime, \( Q_i \) is on the order of \(10^5\) and appears independent of resistivity and impedance. The value aligns with previous results for granular aluminum resonators made using similar processes \cite{grunhauptgralLoss2018}. Since the participation ratio in the compact resonators presented here is larger, this suggests that, within the explored range, surface dielectric losses are not dominant. Recently, in reference \cite{guptaLowLoss2024} it was demonstrated that by optimizing fabrication and using thicker layers at similar resistivities, but lower impedance, quality factors above $10^6$ can be achieved. These low levels of loss make high-impedance grAl resonators appealing for superconducting quantum circuits. At higher power levels, \( Q_i \) saturates to values close to \( 10^6 \), indicating that the dominant loss mechanism, potentially dielectric loss \cite{hunklingerSaturation1972, goldingNonlinearPhonon1973} or quasiparticle bursts \cite{levensoneQuasiparticle2014, gustavssonRelaxation2016}, is saturable, as commonly observed for other superconducting materials.

In Fig.~\ref{fig_resonators_performance}(c) we show that the resonators are comparatively linear, considering their high-impedance. We determine the Kerr coefficient \(K_{11}\), which quantifies the frequency shift of the fundamental mode with the addition of one photon, by performing a linear fit to the frequency shift as a function of \(\bar{n}\) in the high-photon-number regime. Our measurements indicate that \(K_{11}\) is in the range of tens of \si{\hertz}. As can be seen in Fig.~\ref{fig_resonators_performance}(c), resonators with higher resistivity and thinner films exhibit larger \(K_{11}\) values compared to those with lower resistivity and thicker films. This is consistent with modeling grAl as a 1D array of Josephson junctions~\cite{maleevaCircuit2018}, in that case we expect \(K_{11} \propto \rho/V\), where \(V\) is the resonator current mode volume.

An intriguing aspect of our measurements is the observation of a positive frequency shift as a function of photon number in some resonators, as shown in  Fig.~\ref{fig_resonators_performance}(d). We observe that the frequency initially increases at very low photon numbers before the drop corresponding to the Kerr non-linearity discussed in the previous paragraph. Some ring resonators exhibit this effect, while others do not, and we could not identify a correlation of this effect with resistivity, impedance, or other device parameters. Additionally, this positive frequency shift vs. \(\bar{n}\) can vary for the same device in-between cooldowns. Similar observations have recently been reported on resonators integrating a flake of van der Waals cuprate superconductor~\cite{jinExplorings2025}. This behavior is highly relevant for quantum devices operating in the low-power regime, and it needs further investigation to fully understand its underlying mechanisms, possibly related to strongly coupled spurious two-level systems~\cite{Andersson2021Jan, Capelle2020Mar, Kirsh2017Jun}.

In summary, we fabricated and characterized granular aluminum meandered trace ring resonators, achieving high-impedance with a single-step lift-off e-beam lithography process. 
We propose a model in which the impedance and frequency of grAl resonators scale predictably with design parameters, with maximal impedance achieved by minimizing the diameter and correspondingly adjusting the pitch, trace-width and sheet inductance, within material and lithographic constraints.
Our best devices exhibit kinetic inductances up to \SI{4}{\micro\henry} and impedances exceeding \SI{100}{\kilo\ohm}, about 16 times the resistance quantum. 
We demonstrated quality factors on the order of \(10^5\) in the single-photon regime, making them suitable for quantum information processing. 
Their measured self-Kerr non-linearity, in the range of tens of \si{\hertz}, is significantly lower than in Josephson junction arrays. 
Interestingly, we observe an anomalous positive frequency shift at low photon numbers of unknown origin.
Further impedance enhancements could be achieved by reducing the dielectric constant, for example, through backside etching or by detaching from the substrate.

\section*{Acknowledgements}
We are grateful to L. Radtke and S. Diewald for technical assistance. We thank Mathieu Fechant, Horst Hahn, David Niepce, Dennis Rieger, and Patrick Winkel for constructive discussions and feedback. We acknowledge funding from the European Union’s Horizon 2020 research and innovation program under the Marie Skłodowska-Curie grant agreement number 847471 (QUSTEC) and the Federal Ministry of Education and Research (Projects QSolid (FKZ:13N16151) and GeQCoS (FKZ: 13N15683)).  Facilities use was supported by the KIT Nanostructure Service Laboratory (NSL) and by the Karlsruhe Nano Micro Facility (KNMFi). We acknowledge qKit for providing a convenient measurement software framework.

\appendix
\balancecolsandclearpage
\onecolumngrid
\renewcommand{\appendixname}{Appendix}
\section{Setup}
\label{app:setup}

\begin{figure}[H]
\includegraphics[width = 1\columnwidth]{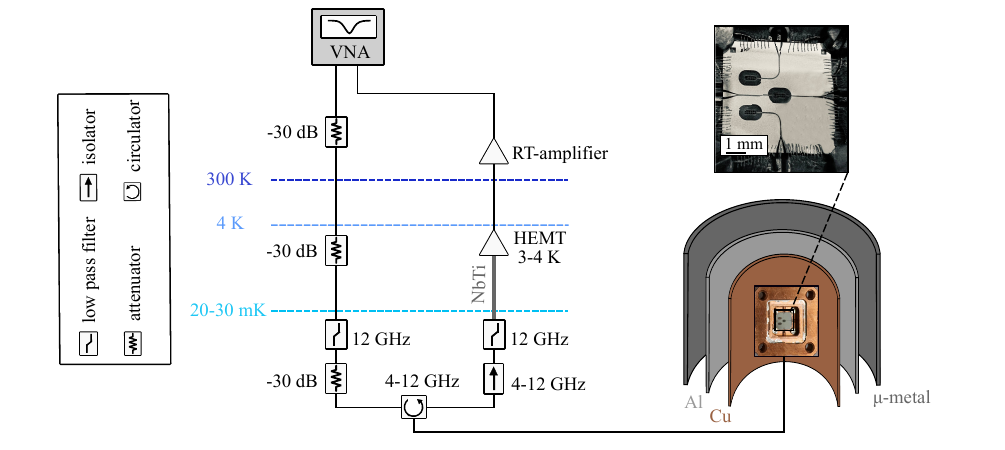}
\caption{\textbf{Experimental Setup.}
 The measurement setup comprises a sample box housing the device under test, which is enclosed within a multilayer magnetic shield to ensure effective isolation. The input signal from a VNA is transmitted through coaxial cables with attenuation applied at each thermal stage of the dilution cryostat before reaching the sample at the 20-30~\si{mK} MXC stage. Reflection measurements are performed by using a cryogenic circulator to separate input and output signals. The reflected signal is subsequently amplified by a HEMT amplifier at the 4K stage and a room-temperature amplifier before being returned to the VNA. Dashed lines in the schematic indicate the thermalization stages and a legend explains the symbols used to represent components such as circulators, isolators, attenuators, and amplifiers in the setup.}
\label{figS2_Setup}
\end{figure}

As shown in Fig.~\ref{figS2_Setup}, each chip's CPW is aluminum wire-bonded to a \SI{50}{\ohm} transmission line, with the ground plane bonded to a non-magnetic, oxygen-free, high-conductivity copper box to minimize spurious modes. This box features a tightly screwed lid to prevent radiation leakage. The entire assembly is mounted on a copper rod attached to the MXC stage of a dilution cryostat and enclosed within a series of magnetic shields similar to the setup in~\cite{grunhauptArgonIonBeam2017}, including a \ce{Cu}/\ce{Al} bilayer and an additional \ce{Mu}-metal shield to ensure effective magnetic shielding.

The input signal, generated by a vector network analyzer (VNA), is transmitted through cables, with attenuators at various cryogenic stages as detailed in Fig.~\ref{figS2_Setup}. For single-port reflection measurements, a cryogenic circulator separates the input and output signals. The reflected output signal then passes through a two-stage isolator, which shields against 4K stage radiation, and through a low-pass filter before reaching a low-noise high electron mobility transistor (HEMT) amplifier via \ce{NbTi} cables with minimal noise contribution. Finally, the amplified signal is sent to commercial room-temperature amplifiers and then back to the VNA output.

\section{Designed and Empirically Extracted Parameters}
\label{app:design_and_extracted_parameters}

We confirmed the accuracy of geometric parameters via SEM imaging. The resonance frequency \(f_0\) was measured experimentally as detailed in App.~\ref{app:setup}. \(L_{\square}\) was determined by fitting the measured \(f_0\) to simulated frequencies based on the nominal design parameters.

The total inductance \(L\) was calculated based on the geometric parameters and \(L_{\square}\), using the formula \(L = l/w \times L_{\square}\), where \(l/w\) represents the number of squares. \(\rho\) was calculated using Eqs.~\ref{eq:kinetic_inductance}, with the critical temperature \(T_\mathrm{c}\) set approximately to 2.2~K for the calculation.

\(Z\) was derived from \(Z = 2 \pi f_0 L\), assuming that the entire ring behaves as an inductor up to the resonance frequency. \(Q_i\) and \(Q_c\) were both extracted using the circle fit method. The value of \(Q_c\) is influenced by the ring's position relative to the feed line, as well as the ring’s radius—the smaller the radius, the weaker the coupling.

A summary of all parameters discussed above is given in Table~\ref{table3_Design}.

\begin{table*}[t]
\renewcommand{\arraystretch}{1.3}
\setlength{\tabcolsep}{4pt}
\caption{Summary of ring resonator parameters, including geometrical dimensions, \(L_{\square}\), \(Z\), and \(f_0\) extracted from both measurements and simulations. \(Q_c\) were determined using the circle fitting method. This table outlines the key design parameters influencing resonator performance, providing a basis for optimization in various applications.}

\vspace{0.2cm}

\begin{tabular}{|p{1.2cm}p{1.2cm}p{1.2cm}p{1.2cm}p{1.2cm}p{1.2cm}p{1.2cm}p{1.2cm}p{1.2cm}p{1.2cm}p{1.2cm}p{1.2cm}|}

\hline
\text{Resonator} & \hspace{5pt}\(r_{\mathrm{in}}\) & \(w\) & \(l\) & \(p\) & \(t\) & \(f_0\) & \(L_{\square}\) & \(L\) & \(C\) & \(Z\) & \hspace{-10pt}\(Q_c \, (\times 10^3)\) \\
 & (\si{\mu\meter}) & (\si{\nano\meter}) & (\si{\mu\meter}) & (\si{\nano\meter}) & (\si{\nano\meter}) & (\si{\giga\hertz}) & (\si{\pico\henry}/\(\square)\) & (\si{\micro\henry}) & (fF) & (\si{\kilo\Omega}) &\\
\hline
1 & 9.1 & 170 & 1849 & 300 & 30 & 5.95 & 180 & 1.95 & 1.46 & 72.9 & 23 \\ \hline
2 & 9.1 & 170 & 1849 & 300 & 30 & 5.98 & 180 & 1.95 & 1.45 & 72.9 & 5 \\ \hline
3 & 7.4 & 150 & 1239 & 300 & 30 & 6.43 & 260 & 2.14 & 1.14 & 86.45 & 5 \\ \hline
4 & 7.4 & 150 & 1239 & 300 & 30 & 6.54 & 260 & 2.14 & 1.10 & 86.45 & 600 \\ \hline
5 & 7.4 & 150 & 1066 & 355 & 30 & 7.02 & 260 & 1.84 & 1.11 & 81.15 & 150 \\ \hline
6 & 7.4 & 150 & 1066 & 355 & 30 & 7.10 & 260 & 1.84 & 1.09 & 81.15 & 20 \\ \hline
7 & 5.2 & 150 & 928 & 200 & 30 & 8.87 & 260 & 1.60 & 0.80 & 89.17 & 25 \\ \hline
8 & 5.2 & 150 & 928 & 200 & 30 & 8.9 & 260 & 1.60 & 0.79 & 89.17 & 22 \\ \hline
9 & 4.8 & 150 & 796 & 200 & 20 & 6.80 & 530 & 2.81 & 0.78 & 120.05 & 50 \\ \hline
10 & 4.8 & 150 & 796 & 200 & 20 & 6.92 & 530 & 2.81 & 0.75 & 120.05 & 30 \\ \hline
11 & 4.8 & 150 & 681 & 240 & 20 & 7.39 & 530 & 2.40 & 0.77 & 111.43 & 230 \\ \hline
12 & 4.8 & 150 & 681 & 240 & 20 & 7.49 & 530 & 2.40 & 0.75 & 111.43 & 130 \\ \hline
13 & 5.6 & 150 & 875 & 300 & 20 & 5.70 & 600 & 3.50 & 0.89 & 125.35 & 200 \\ \hline
14 & 5.6 & 150 & 875 & 300 & 20 & 5.81 & 600 & 3.50 & 0.85 & 125.35 & 130 \\ \hline
15 & 5.6 & 150 & 741 & 253 & 20 & 6.03 & 600 & 2.96 & 0.94 & 112.14 & 60 \\ \hline
16 & 5.6 & 150 & 741 & 253 & 20 & 6.13 & 600 & 2.96 & 0.91 & 112.14 & 6 \\ \hline
17 & 6.7 & 150 & 966 & 325 & 20 & 4.68 & 670 & 4.31 & 1.07 & 126.7 & 140 \\ \hline
18 & 6.7 & 150 & 966 & 325 & 20 & 4.76 & 670 & 4.31 & 1.03 & 126.7 & 50 \\ \hline
19 & 6.7 & 150 & 806 & 400 & 20 & 5.00 & 670 & 3.60 & 1.12 & 113.9 & 8 \\ \hline
20 & 6.7 & 150 & 806 & 400 & 20 & 5.11 & 670 & 3.60 & 1.07 & 113.09 & 6 \\ \hline
21 & 6.7 & 150 & 646 & 518 & 20 & 5.78 & 670 & 2.88 & 1.05 & 104.59 & 260 \\ \hline
22 & 6.7 & 150 & 646 & 518 & 20 & 5.86 & 670 & 2.88 & 1.02 & 104.59 & 30 \\ 

\hline
\end{tabular}
\label{table3_Design}
\end{table*}

\section{Model of Ring Superinductors}
\label{app:model}

\begin{figure}[t]
\includegraphics[width = 1\columnwidth]{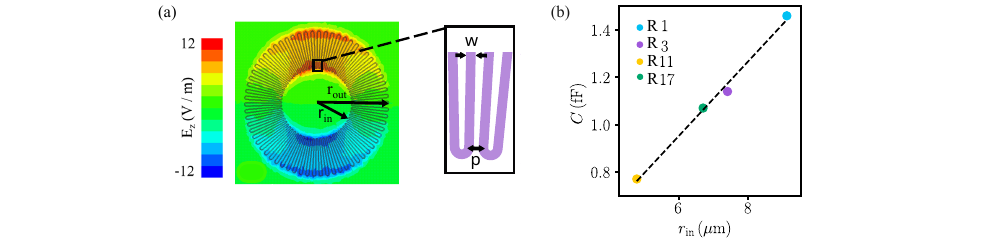}
\caption{\textbf{Geometric Design and Capacitance Scaling in Ring Superinductors.}
(a) Electric field distribution of the ring superinductor, highlighting the capacitor plates and the remaining sections of the ring acting as inductors. We define \(r_{\text{in}}\) and \(r_{\text{out}}\) of the ring resonator as indicated. The zoomed-in plot illustrates how we define \(w\) and \(p\). (b) Capacitance as a function of \(r_{\text{in}}\) for rings with varying inductance per square and resistivity. Each data point corresponds to a specific resonator, with details listed in Table~\ref{table3_Design}. The dashed line acts as a guide to the eye and confirms the linear dependence of capacitance on \(r_{\text{in}}\) for the range of design parameters used here.
}

\label{figS4_RingModel}
\end{figure}

Ignoring the turns, which are approximately equal to the sum of \( \pi r_{\text{in}} \) and \( \pi r_{\text{out}} \), the inductance can be approximated with the expression \( L = N L_{\square} r_{\text{in}}/w \). As we can see in Fig.~\ref{figS4_RingModel}(a), \( w \) is the wire width, \( L_{\square}\) is the inductance per square, and \( r_{\text{in}} \) is the inner radius of the ring. The outer radius \( r_{\text{out}} \) is related to the inner radius as: \( r_{\text{out}} = 2 \, r_{\text{in}}.\) Here, \( N \) is the number of meanders, which is proportional to \(2\pi r_{\text{in}}/p \), where \( p \) is the pitch (the sum of wire width and spacing). Substituting this relationship, we find:

\[
L \propto \frac{r_{\text{in}}^2 L_{\square}}{p w}.
\]

The capacitance \( C \) of the ring resembles that of a coplanar capacitor composed of two square plates with sides \( r_{\text{in}} \) and separated by a length \( 2\,r_{\text{in}} \). The capacitance of such a capacitor scales linearly with the plate width and thus \( r_{\text{in}} \). Therefore we can approximate the ring capacitance by \( C \propto r_{\text{in}}\)\cite{Zypman2019Sep}.

In Fig.~\ref{figS4_RingModel}(b) we confirmed that this linear dependency is consistent with the capacitances determined from HFSS.

The resonance frequency is derived from the two expressions for \(L\) and \(C\) using the harmonic oscillator equation:
\[ f_0 \propto \frac{1}{\sqrt{L C}} \propto \sqrt{\frac{p w}{r_{\text{in}}^3 L_{\square}}}.\]

The impedance \( Z \) depends on both the inductance and capacitance of the ring and can be expressed again using the expressions for \(L\) and \(C\) determined above:
\[
Z \propto \sqrt{ \frac{r_{\text{in}} L_{\square}}{p w} }
\]

 Again, we confirm this proportionality of \( Z \) to the design parameters for the resonators listed in Tab.~\ref{table3_Design} in the plot shown in Fig.~\ref{fig_impedance_model}(a). In addition, we confirm in Fig.~\ref{fig_impedance_model}(b) that the product \( Z f_0 \) only depends on the ring radius. This shows that, for a given fixed frequency, the impedance can be increased (maximizing the product) only by reducing the radius and compensating the associated frequency shift with one of the other design parameters ( \( p, w,~\mathrm{or } ~L_{\square} \)). This compensation does not affect the product \( Z f_0 \).

\section{Fabrication}
\label{app:fabrication}

Device under test includes the ring resonators, CPW, and ground plane, is fabricated on a double-polished, 2~\text{inches} c-plane sapphire wafer with a thickness of \SI{330}{\micro\meter}. The lift-off lithography process (Details in Table~\ref{table1_Fab}) begins with spin coating a PMMA resist layer, followed by the application of a conductive coating. The resonators are patterned using a \SI{50}{kV} e-beam lithography system, followed by the removal of the conductive layer and the development of the resist pattern.

\begin{figure}[H]
\includegraphics[width = 1\columnwidth]{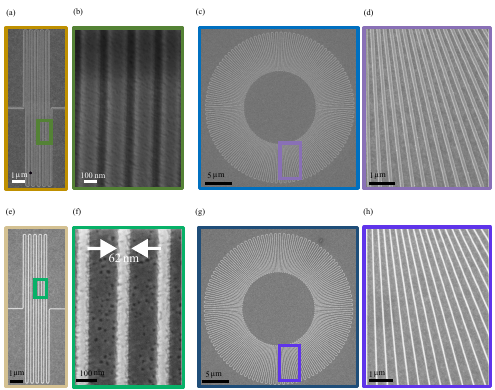}
\caption{\textbf{ Packed and Narrow Meandering grAl Lines.}
SEM images of grAl meandering lines with a thickness of \SI{30}{\nano\meter}, patterned using electron beam lithography.
(a, b) Test structures with meandering lines matching the width of those in ring resonators. These simpler designs were used to optimize the fabrication recipe, achieving a packing density with a pitch of \SI{200}{nm}. 
(c, d) Ring resonators fabricated with the same packing density as the test structures, showing results over longer lines with a greater number of meanders.
(e, f) Test structures highlighting the minimum reproducible line width of about \SI{60}{\nano\meter}. 
(g, h) Ring resonators incorporate the same narrow lines as the test structures, indicating the possibility of making narrow widths in more complex geometries.}
\label{figS1_Fab}
\end{figure}

Subsequently, the wafer is transferred to a UHV Plassys e-beam evaporator, where it undergoes plasma cleaning, an \ce{Ar/O2} descum process. Titanium is then evaporated as a getter material to improve vacuum conditions, with the shutter kept closed during this step. The grAl layer is deposited at room temperature using e-beam evaporation of pure aluminum while flowing a small amount of oxygen through the chamber. After deposition, the grAl film is lifted off by dissolving the resist underneath it. The resistivity of the resulting grAl films ranges from \SI{800}{\micro\ohm\centi\meter} to \SI{2500}{\micro\ohm\centi\meter} as determined by comparing the frequencies extracted from HFSS simulation with the measured ones, SEM images of representative devices are shown in Fig.~\ref{figS1_Fab}.

Fabrication of the CPW starts with resist spin coating, UV lithography patterning, and development. The CPW metallization involves \ce{Al + Nb} deposition using the Plassys evaporator, including \ce{Ar/O2} descum cleaning and \ce{Ti} gettering, followed by lift-off.

After the CPW lift-off process, an additional optical lithography step is performed to deposit the silver ground plane. This step mirrors the CPW fabrication process but includes an additional \ce{Ar} cleaning step to remove any oxide layer on the \ce{Al}, ensuring good galvanic contact between the overlapping silver and \ce{Nb + Al} structures, which is crucial for avoiding parasitic modes. (Fabrication steps in table~\ref{table2_Fab})

After completing the patterning, a protective photoresist layer is applied, and the chip is diced into pieces of \SI{6}{\milli\meter} by \SI{6}{\milli\meter}. The final step involves cleaning off the protective resist layer.

\begin{table}[H]
\centering
\renewcommand{\arraystretch}{1.3} 
\setlength{\tabcolsep}{4pt} 
\caption{ Summary of fabrication details for the ring resonators, including resist spinning, e-beam lithography for patterning, plasma cleaning, and grAl deposition in a dynamic oxygen environment. The resistivity of the resulting grAl films ranges from \SI{800}{\micro\ohm\centi\meter} to \SI{2500}{\micro\ohm\centi\meter}.}
\vspace{0.5cm}
\begin{tabular}{|p{4cm}p{4cm}|} 
\hline
\multicolumn{2}{|c|}{\textbf{Resist Coating}} \\ \hline
Substrate      & C-plane sapphire\\
E-beam resist  & PMMA A4\newline (thickness = \SI{200}{\nano\meter})   \\ 
Acceleration   & \SI{1000}{\rpm\per\s}                             \\ 
Spread cycle   & \SI{300}{\rpm} for \SI{4}{\second}                              \\ 
Spin speed     & \SI{5000}{\rpm} for \SI{60}{\second}                                                               \\
Baking temperature    & \SI{150}{\degreeCelsius} for \SI{3}{\minute}                              \\ 
[1ex]
Conductive coating         & Electra 92 (AR-PC 5090)    \\ 
Acceleration   & \SI{1000}{\rpm\per\s}                             \\ 
Spread cycle   & \SI{500}{\rpm} for \SI{1}{\second}                                                            \\ 
Spin speed     & \SI{2000}{\rpm} for \SI{60}{\second}                             \\ 
Baking temperature    & \SI{85}{\degreeCelsius} for \SI{2}{\minute}                               \\                               
\hline
\multicolumn{2}{|c|}{\textbf{E-beam Exposure}}            \\ \hline
Acceleration voltage      & \SI{50}{\kilo\volt}                       \\ 
Beam current   & \SI{100}{\pico\ampere}                      \\ 
Step size      & \SI{10}{\nano\meter}                                 \\                                     
\hline
\multicolumn{2}{|c|}{\textbf{Development}}             \\ \hline
Conductive coating removal       & Water (RT) for \SI{30}{\second}                                \\                        
[1ex]
Developer      & MIBK 1:3 IPA \newline at \SI{0}{\degreeCelsius} for \SI{30}{\second}               \\                           
\hline
\multicolumn{2}{|c|}{\textbf{Deposition Process}}             \\ \hline
\textbf{Plasma Cleaning}                                       & \\                       
Mass flows                & \ce{O2}/\ce{Ar} (\SI{10}{\sccm}/\SI{5}{\sccm}) \\ 
Beam parameters           & $U_\text{beam} = \SI{200}{\volt}$ \newline $I_\text{beam} = \SI{10}{\milli\ampere}$ \\
[1ex]
\textbf{Gettering} & \\
Ti evaporation rate       & \SI{0.2}{\nano\meter\per\second} for \SI{2}{\minute} \\             
[1ex]
\textbf{grAl Deposition} & \\
Al evaporation rate       & \SI{1}{\nano\meter\per\second}                           \\ 
\ce{O2} pressure                   & \(10^{-5}\) to \(10^{-4}\)~\si{\milli\bar}   \\

\hline
\end{tabular}
\label{table1_Fab}
\end{table}

\begin{table}[H]
\centering
\renewcommand{\arraystretch}{1.3} 
\setlength{\tabcolsep}{4pt} 
\caption{Summary of fabrication details for the CPW and silver ground plane, including resist spinning, UV lithography for patterning, plasma cleaning, and \ce{Nb + Al} metallization. The silver ground plane is deposited using an additional optical lithography step with an \ce{Ar} cleaning stage to remove \ce{Al} oxide, ensuring galvanic contact between overlapping silver and \ce{Nb + Al} structures.}
\vspace{0.5cm}
\begin{tabular}{|p{4cm}p{4cm}|}
\hline
\multicolumn{2}{|c|}{\textbf{Resist Coating}} \\ \hline
Substrate      & C-plane sapphire \\
Photoresist    & AZ 5214E \\ 
Acceleration   & \SI{2000}{\rpm\per\s} \\
Spin speed     & \SI{4000}{\rpm} for \SI{60}{\second} \\
Baking temperature    & \SI{110}{\degreeCelsius} for \SI{50}{\second} \\
\hline
\multicolumn{2}{|c|}{\textbf{UV Exposure}} \\ \hline
\textbf{1st Exposure} & \\
Exposure power & \SI{2}{\milli\watt} for \SI{12}{\second} \\
[1ex]
\textbf{Baking} & \\
Baking temperature    & \SI{120}{\degreeCelsius} for \SI{60}{\second} \\
[1ex]
\textbf{2nd Exposure} & \\
Exposure type  & Flood exposure for \SI{30}{\second} \\
\hline
\multicolumn{2}{|c|}{\textbf{Development}} \\ \hline
Developer      & AZ developer 3:2 water \newline for \SI{30}{\second} \\                           
\hline
\multicolumn{2}{|c|}{\textbf{Deposition Process}} \\ \hline
\textbf{Plasma Cleaning} & \\                       
Mass flows     & \ce{O2}/\ce{Ar} (\SI{10}{\sccm}/\SI{5}{\sccm}) \\ 
Beam parameters & $U_{\text{beam}} = \SI{200}{\volt}$ \newline $I_{\text{beam}} = \SI{10}{\milli\ampere}$ \\ 
[1ex]
\textbf{Gettering} & \\
\ce{Ti} evaporation rate & \SI{0.2}{\nano\meter\per\second} for \SI{2}{\minute} \\
[1ex]
\textbf{\ce{Nb} + \ce{Al} Deposition} & \\
\ce{Nb} evaporation rate & \SI{1}{\nano\meter\per\second} \\ 
Al evaporation rate & \SI{1}{\nano\meter\per\second} \\ 
\hline
\textbf{\ce{Ag} Deposition} & \\
\ce{Ag} evaporation rate & \SI{1}{\nano\meter\per\second} \\ 
\hline
\end{tabular}
\label{table2_Fab}
\end{table}

\section{Numerical Simulation of High-Resistivity Ring Resonators}
\label{app:Numerical_Sim}

To explore the higher resistivity regime of grAl below the superconducting-to-insulator phase transition (SIT), we simulated grAl ring resonators with \(L_\square\) ranging from \SI{1200}{\pico\henry/\text{sq}} to \SI{1800}{\pico\henry/\text{sq}} which could be achieved by using \(\rho\) from \SI{4338}{\micro\ohm\centi\meter} to \SI{6490}{\micro\ohm\centi\meter}. These simulations are performed using the eigenmode solver in ANSYS HFSS to determine intrinsic resonant frequencies.

\begin{figure}[H]
\includegraphics[width=\columnwidth]{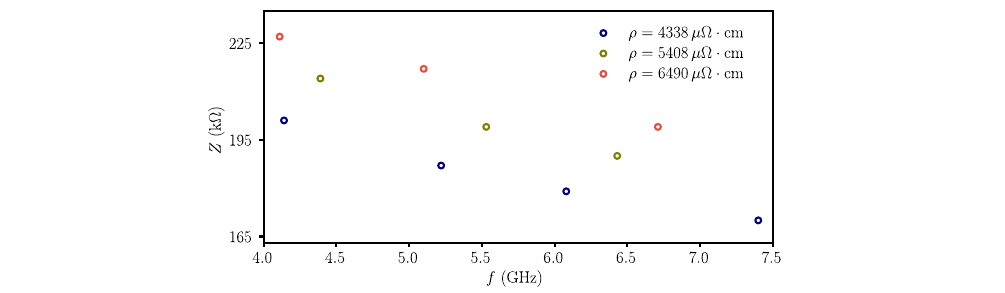}
\caption{\textbf{Prospective grAl Impedance Optimization.} Simulated impedances up to the fundamental mode, indicated by the x-axis, for the designs summarized in Table~\ref{table4_Design}.
}
\label{figS8_Simulated_impedance_frequency}
\end{figure}

 The simulated parameters, including resonator dimensions, inductance, resonance frequencies, and corresponding impedances, are presented in Table~\ref{table4_Design} and the impedance values are reported in Fig.~\ref{figS8_Simulated_impedance_frequency}, providing a perspective for future impedance increases in grAl resonators. 

\begin{table*}[t]
\vspace*{-0.55cm}
\centering
\renewcommand{\arraystretch}{1.3}
\setlength{\tabcolsep}{4pt}
\caption{Summary of simulated ring resonator parameters, including geometrical dimensions, \(L_{\square}\), \(L\), \(Z\), and \(f_0\). The parameters were derived using eigenmode simulations in ANSYS HFSS, focusing on the higher resistivity regime of grAl below SIT. These results provide guidelines for designing resonators within the \SIrange{4}{8}{\giga\hertz} frequency range by optimizing impedance and frequency characteristics.}
\vspace{0.2cm}

\begin{tabular}{|p{1.35cm}p{1.35cm}p{1.35cm}p{1.35cm}p{1.35cm}p{1.35cm}p{1.35cm}p{1.35cm}p{1.35cm}p{1.35cm}p{1.35cm}|}
\hline
\text{\(\text{Resonator}\)} & \(r_{\mathrm{in}}\) & \(w\) & \(l\) & \(p\) & \(t\) & \(f_0\) & \(L_{\square}\) & \(L\) & \(C\) & \(Z\)  \\
 & (\si{\mu\meter}) & (\si{\nano\meter}) & (\si{\mu\meter}) & (\si{\nano\meter}) & (\si{\nano\meter}) & (\si{\giga\hertz}) & (\si{\pico\henry}/\(\square\)) & (\si{\mu\henry}) & (fF) & (\si{\kilo\Omega}) \\
 
 \hline
23 & 3.2 & 120 & 367 & 200 & 20 & 7.3 & 1200 & 3.67 & 0.518 & 170  \\\hline
24 & 3.69 & 120 & 469.8 & 200 & 20 & 6.07 & 1200 & 4.69 & 0.586 & 179 \\\hline
25 & 4.07 & 120 &  571.9 & 200 & 20 & 5.22 & 1200 & 5.71 & 0.651 & 187 \\\hline
26 & 4.77 & 120 & 775.7 & 200 & 20 & 4.14 & 1200 & 7.75 & 0.763 & 201 \\\hline
27 & 3.31 & 120 & 377 & 200 & 20 & 6.43 & 1500 & 4.7 & 0.521 & 190 \\\hline
28 & 3.62 & 120 & 459 & 200 & 20 & 5.53 & 1500 & 5.7 & 0.581 & 199 \\\hline
29 & 4.2 & 120 & 622 & 200 & 20 & 4.39 & 1500 & 7.7 & 0.683 & 214 \\\hline
30 & 2.9 & 120 & 316 & 200 & 20 & 6.7 & 1800 & 4.7 & 0.480 & 199 \\\hline
31 & 3.6 & 120 & 452 & 200 & 20 & 5.09 & 1800 & 6.7 & 0.584 & 217 \\\hline
32 & 4.13 & 120 & 588 & 200 & 20 & 4.1 & 1800 & 8.8 & 0.685 & 227 \\
\hline
\end{tabular}
\label{table4_Design}
\end{table*}

\twocolumngrid
\bibliography{references}

\begin{thebibliography}{70}%
\makeatletter
\providecommand \@ifxundefined [1]{%
 \@ifx{#1\undefined}
}%
\providecommand \@ifnum [1]{%
 \ifnum #1\expandafter \@firstoftwo
 \else \expandafter \@secondoftwo
 \fi
}%
\providecommand \@ifx [1]{%
 \ifx #1\expandafter \@firstoftwo
 \else \expandafter \@secondoftwo
 \fi
}%
\providecommand \natexlab [1]{#1}%
\providecommand \enquote  [1]{``#1''}%
\providecommand \bibnamefont  [1]{#1}%
\providecommand \bibfnamefont [1]{#1}%
\providecommand \citenamefont [1]{#1}%
\providecommand \href@noop [0]{\@secondoftwo}%
\providecommand \href [0]{\begingroup \@sanitize@url \@href}%
\providecommand \@href[1]{\@@startlink{#1}\@@href}%
\providecommand \@@href[1]{\endgroup#1\@@endlink}%
\providecommand \@sanitize@url [0]{\catcode `\\12\catcode `\$12\catcode `\&12\catcode `\#12\catcode `\^12\catcode `\_12\catcode `\%12\relax}%
\providecommand \@@startlink[1]{}%
\providecommand \@@endlink[0]{}%
\providecommand \url  [0]{\begingroup\@sanitize@url \@url }%
\providecommand \@url [1]{\endgroup\@href {#1}{\urlprefix }}%
\providecommand \urlprefix  [0]{URL }%
\providecommand \Eprint [0]{\href }%
\providecommand \doibase [0]{https://doi.org/}%
\providecommand \selectlanguage [0]{\@gobble}%
\providecommand \bibinfo  [0]{\@secondoftwo}%
\providecommand \bibfield  [0]{\@secondoftwo}%
\providecommand \translation [1]{[#1]}%
\providecommand \BibitemOpen [0]{}%
\providecommand \bibitemStop [0]{}%
\providecommand \bibitemNoStop [0]{.\EOS\space}%
\providecommand \EOS [0]{\spacefactor3000\relax}%
\providecommand \BibitemShut  [1]{\csname bibitem#1\endcsname}%
\let\auto@bib@innerbib\@empty
\bibitem [{\citenamefont {Devoret}\ \emph {et~al.}(2007)\citenamefont {Devoret}, \citenamefont {Girvin},\ and\ \citenamefont {Schoelkopf}}]{devoretCircuitQED2007}%
  \BibitemOpen
  \bibfield  {author} {\bibinfo {author} {\bibfnamefont {M.}~\bibnamefont {Devoret}}, \bibinfo {author} {\bibfnamefont {S.}~\bibnamefont {Girvin}},\ and\ \bibinfo {author} {\bibfnamefont {R.}~\bibnamefont {Schoelkopf}},\ }\bibfield  {title} {\bibinfo {title} {Circuit-{QED}: How strong can the coupling between a josephson junction atom and a transmission line resonator {B}e?},\ }\href {https://doi.org/10.1002/andp.200751910-1109} {\bibfield  {journal} {\bibinfo  {journal} {Ann. Phys.}\ }\textbf {\bibinfo {volume} {519}},\ \bibinfo {pages} {767} (\bibinfo {year} {2007})}\BibitemShut {NoStop}%
\bibitem [{\citenamefont {Burkard}\ \emph {et~al.}(2023)\citenamefont {Burkard}, \citenamefont {Ladd}, \citenamefont {Pan}, \citenamefont {Nichol},\ and\ \citenamefont {Petta}}]{burkardSpinQubits2023}%
  \BibitemOpen
  \bibfield  {author} {\bibinfo {author} {\bibfnamefont {G.}~\bibnamefont {Burkard}}, \bibinfo {author} {\bibfnamefont {T.~D.}\ \bibnamefont {Ladd}}, \bibinfo {author} {\bibfnamefont {A.}~\bibnamefont {Pan}}, \bibinfo {author} {\bibfnamefont {J.~M.}\ \bibnamefont {Nichol}},\ and\ \bibinfo {author} {\bibfnamefont {J.~R.}\ \bibnamefont {Petta}},\ }\bibfield  {title} {\bibinfo {title} {Semiconductor spin qubits},\ }\href {https://doi.org/10.1103/RevModPhys.95.025003} {\bibfield  {journal} {\bibinfo  {journal} {Rev. Mod. Phys.}\ }\textbf {\bibinfo {volume} {95}},\ \bibinfo {pages} {025003} (\bibinfo {year} {2023})}\BibitemShut {NoStop}%
\bibitem [{\citenamefont {Arrangoiz-Arriola}\ \emph {et~al.}(2018)\citenamefont {Arrangoiz-Arriola}, \citenamefont {Wollack}, \citenamefont {Pechal}, \citenamefont {Witmer}, \citenamefont {Hill},\ and\ \citenamefont {Safavi-Naeini}}]{arrangoiz-arriolaCoupling2018}%
  \BibitemOpen
  \bibfield  {author} {\bibinfo {author} {\bibfnamefont {P.}~\bibnamefont {Arrangoiz-Arriola}}, \bibinfo {author} {\bibfnamefont {E.~A.}\ \bibnamefont {Wollack}}, \bibinfo {author} {\bibfnamefont {M.}~\bibnamefont {Pechal}}, \bibinfo {author} {\bibfnamefont {J.~D.}\ \bibnamefont {Witmer}}, \bibinfo {author} {\bibfnamefont {J.~T.}\ \bibnamefont {Hill}},\ and\ \bibinfo {author} {\bibfnamefont {A.~H.}\ \bibnamefont {Safavi-Naeini}},\ }\bibfield  {title} {\bibinfo {title} {Coupling a superconducting quantum circuit to a phononic crystal defect cavity},\ }\href {https://doi.org/10.1103/PhysRevX.8.031007} {\bibfield  {journal} {\bibinfo  {journal} {Phys. Rev. X}\ }\textbf {\bibinfo {volume} {8}},\ \bibinfo {pages} {031007} (\bibinfo {year} {2018})}\BibitemShut {NoStop}%
\bibitem [{\citenamefont {Bozkurt}\ \emph {et~al.}(2023)\citenamefont {Bozkurt}, \citenamefont {Zhao}, \citenamefont {Joshi}, \citenamefont {LeDuc}, \citenamefont {Day},\ and\ \citenamefont {Mirhosseini}}]{bozkurtQuantum2023}%
  \BibitemOpen
  \bibfield  {author} {\bibinfo {author} {\bibfnamefont {A.}~\bibnamefont {Bozkurt}}, \bibinfo {author} {\bibfnamefont {H.}~\bibnamefont {Zhao}}, \bibinfo {author} {\bibfnamefont {C.}~\bibnamefont {Joshi}}, \bibinfo {author} {\bibfnamefont {H.~G.}\ \bibnamefont {LeDuc}}, \bibinfo {author} {\bibfnamefont {P.~K.}\ \bibnamefont {Day}},\ and\ \bibinfo {author} {\bibfnamefont {M.}~\bibnamefont {Mirhosseini}},\ }\bibfield  {title} {\bibinfo {title} {A quantum electromechanical interface for long-lived phonons},\ }\href {https://doi.org/10.1038/s41567-023-02080-w} {\bibfield  {journal} {\bibinfo  {journal} {Nat. Phys.}\ }\textbf {\bibinfo {volume} {19}},\ \bibinfo {pages} {1326} (\bibinfo {year} {2023})}\BibitemShut {NoStop}%
\bibitem [{\citenamefont {André}\ \emph {et~al.}(2006)\citenamefont {André}, \citenamefont {DeMille}, \citenamefont {Doyle}, \citenamefont {Lukin}, \citenamefont {Maxwell}, \citenamefont {Rabl}, \citenamefont {Schoelkopf},\ and\ \citenamefont {Zoller}}]{andreCoherent2006}%
  \BibitemOpen
  \bibfield  {author} {\bibinfo {author} {\bibfnamefont {A.}~\bibnamefont {André}}, \bibinfo {author} {\bibfnamefont {D.}~\bibnamefont {DeMille}}, \bibinfo {author} {\bibfnamefont {J.~M.}\ \bibnamefont {Doyle}}, \bibinfo {author} {\bibfnamefont {M.~D.}\ \bibnamefont {Lukin}}, \bibinfo {author} {\bibfnamefont {S.~E.}\ \bibnamefont {Maxwell}}, \bibinfo {author} {\bibfnamefont {P.}~\bibnamefont {Rabl}}, \bibinfo {author} {\bibfnamefont {R.~J.}\ \bibnamefont {Schoelkopf}},\ and\ \bibinfo {author} {\bibfnamefont {P.}~\bibnamefont {Zoller}},\ }\bibfield  {title} {\bibinfo {title} {A coherent all-electrical interface between polar molecules and mesoscopic superconducting resonators},\ }\href {https://doi.org/10.1038/nphys386} {\bibfield  {journal} {\bibinfo  {journal} {Nat. Phys.}\ }\textbf {\bibinfo {volume} {2}},\ \bibinfo {pages} {636} (\bibinfo {year} {2006})}\BibitemShut {NoStop}%
\bibitem [{\citenamefont {Manucharyan}\ \emph {et~al.}(2012)\citenamefont {Manucharyan}, \citenamefont {Masluk}, \citenamefont {Kamal}, \citenamefont {Koch}, \citenamefont {Glazman},\ and\ \citenamefont {Devoret}}]{manucharyanSuperinductor2012}%
  \BibitemOpen
  \bibfield  {author} {\bibinfo {author} {\bibfnamefont {V.~E.}\ \bibnamefont {Manucharyan}}, \bibinfo {author} {\bibfnamefont {N.~A.}\ \bibnamefont {Masluk}}, \bibinfo {author} {\bibfnamefont {A.}~\bibnamefont {Kamal}}, \bibinfo {author} {\bibfnamefont {J.}~\bibnamefont {Koch}}, \bibinfo {author} {\bibfnamefont {L.~I.}\ \bibnamefont {Glazman}},\ and\ \bibinfo {author} {\bibfnamefont {M.~H.}\ \bibnamefont {Devoret}},\ }\bibfield  {title} {\bibinfo {title} {Evidence for coherent quantum phase slips across a {{Josephson}} junction array},\ }\href {https://doi.org/10.1103/PhysRevB.85.024521} {\bibfield  {journal} {\bibinfo  {journal} {Phys. Rev. B}\ }\textbf {\bibinfo {volume} {85}},\ \bibinfo {pages} {024521} (\bibinfo {year} {2012})}\BibitemShut {NoStop}%
\bibitem [{\citenamefont {Masluk}\ \emph {et~al.}(2012)\citenamefont {Masluk}, \citenamefont {Pop}, \citenamefont {Kamal}, \citenamefont {Minev},\ and\ \citenamefont {Devoret}}]{maslukMicrowave2012}%
  \BibitemOpen
  \bibfield  {author} {\bibinfo {author} {\bibfnamefont {N.~A.}\ \bibnamefont {Masluk}}, \bibinfo {author} {\bibfnamefont {I.~M.}\ \bibnamefont {Pop}}, \bibinfo {author} {\bibfnamefont {A.}~\bibnamefont {Kamal}}, \bibinfo {author} {\bibfnamefont {Z.~K.}\ \bibnamefont {Minev}},\ and\ \bibinfo {author} {\bibfnamefont {M.~H.}\ \bibnamefont {Devoret}},\ }\bibfield  {title} {\bibinfo {title} {Microwave characterization of josephson junction arrays: Implementing a low loss superinductance},\ }\href {https://doi.org/10.1103/PhysRevLett.109.137002} {\bibfield  {journal} {\bibinfo  {journal} {Phys. Rev. Lett.}\ }\textbf {\bibinfo {volume} {109}},\ \bibinfo {pages} {137002} (\bibinfo {year} {2012})}\BibitemShut {NoStop}%
\bibitem [{\citenamefont {Manucharyan}\ \emph {et~al.}(2009)\citenamefont {Manucharyan}, \citenamefont {Koch}, \citenamefont {Glazman},\ and\ \citenamefont {Devoret}}]{manucharyanFluxonium2009}%
  \BibitemOpen
  \bibfield  {author} {\bibinfo {author} {\bibfnamefont {V.~E.}\ \bibnamefont {Manucharyan}}, \bibinfo {author} {\bibfnamefont {J.}~\bibnamefont {Koch}}, \bibinfo {author} {\bibfnamefont {L.~I.}\ \bibnamefont {Glazman}},\ and\ \bibinfo {author} {\bibfnamefont {M.~H.}\ \bibnamefont {Devoret}},\ }\bibfield  {title} {\bibinfo {title} {Fluxonium: Single cooper-pair circuit free of charge offsets},\ }\href {https://doi.org/10.1126/science.1175552} {\bibfield  {journal} {\bibinfo  {journal} {Science}\ }\textbf {\bibinfo {volume} {326}},\ \bibinfo {pages} {113} (\bibinfo {year} {2009})}\BibitemShut {NoStop}%
\bibitem [{\citenamefont {Pop}\ \emph {et~al.}(2014)\citenamefont {Pop}, \citenamefont {Geerlings}, \citenamefont {Catelani}, \citenamefont {Schoelkopf}, \citenamefont {Glazman},\ and\ \citenamefont {Devoret}}]{popCoherent2014}%
  \BibitemOpen
  \bibfield  {author} {\bibinfo {author} {\bibfnamefont {I.~M.}\ \bibnamefont {Pop}}, \bibinfo {author} {\bibfnamefont {K.}~\bibnamefont {Geerlings}}, \bibinfo {author} {\bibfnamefont {G.}~\bibnamefont {Catelani}}, \bibinfo {author} {\bibfnamefont {R.~J.}\ \bibnamefont {Schoelkopf}}, \bibinfo {author} {\bibfnamefont {L.~I.}\ \bibnamefont {Glazman}},\ and\ \bibinfo {author} {\bibfnamefont {M.~H.}\ \bibnamefont {Devoret}},\ }\bibfield  {title} {\bibinfo {title} {Coherent suppression of electromagnetic dissipation due to superconducting quasiparticles},\ }\href {https://doi.org/10.1038/nature13017} {\bibfield  {journal} {\bibinfo  {journal} {Nature}\ }\textbf {\bibinfo {volume} {508}},\ \bibinfo {pages} {369} (\bibinfo {year} {2014})}\BibitemShut {NoStop}%
\bibitem [{\citenamefont {Grünhaupt}\ \emph {et~al.}(2019)\citenamefont {Grünhaupt}, \citenamefont {Spiecker}, \citenamefont {Gusenkova}, \citenamefont {Maleeva}, \citenamefont {Skacel}, \citenamefont {Takmakov}, \citenamefont {Valenti}, \citenamefont {Winkel}, \citenamefont {Rotzinger}, \citenamefont {Wernsdorfer}, \citenamefont {Ustinov},\ and\ \citenamefont {Pop}}]{grunhauptGranular2019}%
  \BibitemOpen
  \bibfield  {author} {\bibinfo {author} {\bibfnamefont {L.}~\bibnamefont {Grünhaupt}}, \bibinfo {author} {\bibfnamefont {M.}~\bibnamefont {Spiecker}}, \bibinfo {author} {\bibfnamefont {D.}~\bibnamefont {Gusenkova}}, \bibinfo {author} {\bibfnamefont {N.}~\bibnamefont {Maleeva}}, \bibinfo {author} {\bibfnamefont {S.~T.}\ \bibnamefont {Skacel}}, \bibinfo {author} {\bibfnamefont {I.}~\bibnamefont {Takmakov}}, \bibinfo {author} {\bibfnamefont {F.}~\bibnamefont {Valenti}}, \bibinfo {author} {\bibfnamefont {P.}~\bibnamefont {Winkel}}, \bibinfo {author} {\bibfnamefont {H.}~\bibnamefont {Rotzinger}}, \bibinfo {author} {\bibfnamefont {W.}~\bibnamefont {Wernsdorfer}}, \bibinfo {author} {\bibfnamefont {A.~V.}\ \bibnamefont {Ustinov}},\ and\ \bibinfo {author} {\bibfnamefont {I.~M.}\ \bibnamefont {Pop}},\ }\bibfield  {title} {\bibinfo {title} {Granular aluminium as a superconducting material for high-impedance quantum circuits},\ }\href {https://doi.org/10.1038/s41563-019-0350-3} {\bibfield  {journal} {\bibinfo
  {journal} {Nat. Mater.}\ }\textbf {\bibinfo {volume} {18}},\ \bibinfo {pages} {816} (\bibinfo {year} {2019})}\BibitemShut {NoStop}%
\bibitem [{\citenamefont {{Pita-Vidal}}\ \emph {et~al.}(2020)\citenamefont {{Pita-Vidal}}, \citenamefont {Bargerbos}, \citenamefont {Yang}, \citenamefont {Van~Woerkom}, \citenamefont {Pfaff}, \citenamefont {Haider}, \citenamefont {Krogstrup}, \citenamefont {Kouwenhoven}, \citenamefont {De~Lange},\ and\ \citenamefont {Kou}}]{pita-vidalFluxonium2020}%
  \BibitemOpen
  \bibfield  {author} {\bibinfo {author} {\bibfnamefont {M.}~\bibnamefont {{Pita-Vidal}}}, \bibinfo {author} {\bibfnamefont {A.}~\bibnamefont {Bargerbos}}, \bibinfo {author} {\bibfnamefont {C.-K.}\ \bibnamefont {Yang}}, \bibinfo {author} {\bibfnamefont {D.~J.}\ \bibnamefont {Van~Woerkom}}, \bibinfo {author} {\bibfnamefont {W.}~\bibnamefont {Pfaff}}, \bibinfo {author} {\bibfnamefont {N.}~\bibnamefont {Haider}}, \bibinfo {author} {\bibfnamefont {P.}~\bibnamefont {Krogstrup}}, \bibinfo {author} {\bibfnamefont {L.~P.}\ \bibnamefont {Kouwenhoven}}, \bibinfo {author} {\bibfnamefont {G.}~\bibnamefont {De~Lange}},\ and\ \bibinfo {author} {\bibfnamefont {A.}~\bibnamefont {Kou}},\ }\bibfield  {title} {\bibinfo {title} {Gate-tunable field-compatible fluxonium},\ }\href {https://doi.org/10.1103/PhysRevApplied.14.064038} {\bibfield  {journal} {\bibinfo  {journal} {Phys. Rev. Appl.}\ }\textbf {\bibinfo {volume} {14}},\ \bibinfo {pages} {064038} (\bibinfo {year} {2020})}\BibitemShut {NoStop}%
\bibitem [{\citenamefont {Nguyen}\ \emph {et~al.}(2022)\citenamefont {Nguyen}, \citenamefont {Koolstra}, \citenamefont {Kim}, \citenamefont {Morvan}, \citenamefont {Chistolini}, \citenamefont {Singh}, \citenamefont {Nesterov}, \citenamefont {J{\"u}nger}, \citenamefont {Chen}, \citenamefont {Pedramrazi}, \citenamefont {Mitchell}, \citenamefont {Kreikebaum}, \citenamefont {Puri}, \citenamefont {Santiago},\ and\ \citenamefont {Siddiqi}}]{nguyenFluxonium2022}%
  \BibitemOpen
  \bibfield  {author} {\bibinfo {author} {\bibfnamefont {L.~B.}\ \bibnamefont {Nguyen}}, \bibinfo {author} {\bibfnamefont {G.}~\bibnamefont {Koolstra}}, \bibinfo {author} {\bibfnamefont {Y.}~\bibnamefont {Kim}}, \bibinfo {author} {\bibfnamefont {A.}~\bibnamefont {Morvan}}, \bibinfo {author} {\bibfnamefont {T.}~\bibnamefont {Chistolini}}, \bibinfo {author} {\bibfnamefont {S.}~\bibnamefont {Singh}}, \bibinfo {author} {\bibfnamefont {K.~N.}\ \bibnamefont {Nesterov}}, \bibinfo {author} {\bibfnamefont {C.}~\bibnamefont {J{\"u}nger}}, \bibinfo {author} {\bibfnamefont {L.}~\bibnamefont {Chen}}, \bibinfo {author} {\bibfnamefont {Z.}~\bibnamefont {Pedramrazi}}, \bibinfo {author} {\bibfnamefont {B.~K.}\ \bibnamefont {Mitchell}}, \bibinfo {author} {\bibfnamefont {J.~M.}\ \bibnamefont {Kreikebaum}}, \bibinfo {author} {\bibfnamefont {S.}~\bibnamefont {Puri}}, \bibinfo {author} {\bibfnamefont {D.~I.}\ \bibnamefont {Santiago}},\ and\ \bibinfo {author} {\bibfnamefont {I.}~\bibnamefont {Siddiqi}},\ }\bibfield  {title}
  {\bibinfo {title} {Blueprint for a high-performance fluxonium quantum processor},\ }\href {https://doi.org/10.1103/PRXQuantum.3.037001} {\bibfield  {journal} {\bibinfo  {journal} {PRX Quantum}\ }\textbf {\bibinfo {volume} {3}},\ \bibinfo {pages} {037001} (\bibinfo {year} {2022})}\BibitemShut {NoStop}%
\bibitem [{\citenamefont {Peruzzo}\ \emph {et~al.}(2021)\citenamefont {Peruzzo}, \citenamefont {Hassani}, \citenamefont {Szep}, \citenamefont {Trioni}, \citenamefont {Redchenko}, \citenamefont {Žemlička},\ and\ \citenamefont {Fink}}]{peruzzoGeoQubits2021}%
  \BibitemOpen
  \bibfield  {author} {\bibinfo {author} {\bibfnamefont {M.}~\bibnamefont {Peruzzo}}, \bibinfo {author} {\bibfnamefont {F.}~\bibnamefont {Hassani}}, \bibinfo {author} {\bibfnamefont {G.}~\bibnamefont {Szep}}, \bibinfo {author} {\bibfnamefont {A.}~\bibnamefont {Trioni}}, \bibinfo {author} {\bibfnamefont {E.}~\bibnamefont {Redchenko}}, \bibinfo {author} {\bibfnamefont {M.}~\bibnamefont {Žemlička}},\ and\ \bibinfo {author} {\bibfnamefont {J.~M.}\ \bibnamefont {Fink}},\ }\bibfield  {title} {\bibinfo {title} {Geometric superinductance qubits: Controlling phase delocalization across a single josephson junction},\ }\href {https://doi.org/10.1103/PRXQuantum.2.040341} {\bibfield  {journal} {\bibinfo  {journal} {PRX Quantum}\ }\textbf {\bibinfo {volume} {2}},\ \bibinfo {pages} {040341} (\bibinfo {year} {2021})}\BibitemShut {NoStop}%
\bibitem [{\citenamefont {Kalashnikov}\ \emph {et~al.}(2020)\citenamefont {Kalashnikov}, \citenamefont {Hsieh}, \citenamefont {Zhang}, \citenamefont {Lu}, \citenamefont {Kamenov}, \citenamefont {Di~Paolo}, \citenamefont {Blais}, \citenamefont {Gershenson},\ and\ \citenamefont {Bell}}]{kalashnikovBifluxon2020}%
  \BibitemOpen
  \bibfield  {author} {\bibinfo {author} {\bibfnamefont {K.}~\bibnamefont {Kalashnikov}}, \bibinfo {author} {\bibfnamefont {W.~T.}\ \bibnamefont {Hsieh}}, \bibinfo {author} {\bibfnamefont {W.}~\bibnamefont {Zhang}}, \bibinfo {author} {\bibfnamefont {W.-S.}\ \bibnamefont {Lu}}, \bibinfo {author} {\bibfnamefont {P.}~\bibnamefont {Kamenov}}, \bibinfo {author} {\bibfnamefont {A.}~\bibnamefont {Di~Paolo}}, \bibinfo {author} {\bibfnamefont {A.}~\bibnamefont {Blais}}, \bibinfo {author} {\bibfnamefont {M.~E.}\ \bibnamefont {Gershenson}},\ and\ \bibinfo {author} {\bibfnamefont {M.}~\bibnamefont {Bell}},\ }\bibfield  {title} {\bibinfo {title} {Bifluxon: Fluxon-parity-protected superconducting qubit},\ }\href {https://doi.org/10.1103/PRXQuantum.1.010307} {\bibfield  {journal} {\bibinfo  {journal} {PRX Quantum}\ }\textbf {\bibinfo {volume} {1}},\ \bibinfo {pages} {010307} (\bibinfo {year} {2020})}\BibitemShut {NoStop}%
\bibitem [{\citenamefont {Brooks}\ \emph {et~al.}(2013)\citenamefont {Brooks}, \citenamefont {Kitaev},\ and\ \citenamefont {Preskill}}]{brooks0-pi2013}%
  \BibitemOpen
  \bibfield  {author} {\bibinfo {author} {\bibfnamefont {P.}~\bibnamefont {Brooks}}, \bibinfo {author} {\bibfnamefont {A.}~\bibnamefont {Kitaev}},\ and\ \bibinfo {author} {\bibfnamefont {J.}~\bibnamefont {Preskill}},\ }\bibfield  {title} {\bibinfo {title} {Protected gates for superconducting qubits},\ }\href {https://doi.org/10.1103/PhysRevA.87.052306} {\bibfield  {journal} {\bibinfo  {journal} {Phys. Rev. A}\ }\textbf {\bibinfo {volume} {87}},\ \bibinfo {pages} {052306} (\bibinfo {year} {2013})}\BibitemShut {NoStop}%
\bibitem [{\citenamefont {Groszkowski}\ \emph {et~al.}(2018)\citenamefont {Groszkowski}, \citenamefont {Paolo}, \citenamefont {Grimsmo}, \citenamefont {Blais}, \citenamefont {Schuster}, \citenamefont {Houck},\ and\ \citenamefont {Koch}}]{groszkowski0pi02018}%
  \BibitemOpen
  \bibfield  {author} {\bibinfo {author} {\bibfnamefont {P.}~\bibnamefont {Groszkowski}}, \bibinfo {author} {\bibfnamefont {A.~D.}\ \bibnamefont {Paolo}}, \bibinfo {author} {\bibfnamefont {A.~L.}\ \bibnamefont {Grimsmo}}, \bibinfo {author} {\bibfnamefont {A.}~\bibnamefont {Blais}}, \bibinfo {author} {\bibfnamefont {D.~I.}\ \bibnamefont {Schuster}}, \bibinfo {author} {\bibfnamefont {A.~A.}\ \bibnamefont {Houck}},\ and\ \bibinfo {author} {\bibfnamefont {J.}~\bibnamefont {Koch}},\ }\bibfield  {title} {\bibinfo {title} {Coherence properties of the 0- {\emph{{$\pi$}}} qubit},\ }\href {https://doi.org/10.1088/1367-2630/aab7cd} {\bibfield  {journal} {\bibinfo  {journal} {New J. Phys.}\ }\textbf {\bibinfo {volume} {20}},\ \bibinfo {pages} {043053} (\bibinfo {year} {2018})}\BibitemShut {NoStop}%
\bibitem [{\citenamefont {Gyenis}\ \emph {et~al.}(2021)\citenamefont {Gyenis}, \citenamefont {Mundada}, \citenamefont {Di~Paolo}, \citenamefont {Hazard}, \citenamefont {You}, \citenamefont {Schuster}, \citenamefont {Koch}, \citenamefont {Blais},\ and\ \citenamefont {Houck}}]{gyenis0pi2021}%
  \BibitemOpen
  \bibfield  {author} {\bibinfo {author} {\bibfnamefont {A.}~\bibnamefont {Gyenis}}, \bibinfo {author} {\bibfnamefont {P.~S.}\ \bibnamefont {Mundada}}, \bibinfo {author} {\bibfnamefont {A.}~\bibnamefont {Di~Paolo}}, \bibinfo {author} {\bibfnamefont {T.~M.}\ \bibnamefont {Hazard}}, \bibinfo {author} {\bibfnamefont {X.}~\bibnamefont {You}}, \bibinfo {author} {\bibfnamefont {D.~I.}\ \bibnamefont {Schuster}}, \bibinfo {author} {\bibfnamefont {J.}~\bibnamefont {Koch}}, \bibinfo {author} {\bibfnamefont {A.}~\bibnamefont {Blais}},\ and\ \bibinfo {author} {\bibfnamefont {A.~A.}\ \bibnamefont {Houck}},\ }\bibfield  {title} {\bibinfo {title} {Experimental realization of a protected superconducting circuit derived from the 0 -- {$\pi$} qubit},\ }\href {https://doi.org/10.1103/PRXQuantum.2.010339} {\bibfield  {journal} {\bibinfo  {journal} {PRX Quantum}\ }\textbf {\bibinfo {volume} {2}},\ \bibinfo {pages} {010339} (\bibinfo {year} {2021})}\BibitemShut {NoStop}%
\bibitem [{\citenamefont {Shaikhaidarov}\ \emph {et~al.}(2022)\citenamefont {Shaikhaidarov}, \citenamefont {Kim}, \citenamefont {Dunstan}, \citenamefont {Antonov}, \citenamefont {Linzen}, \citenamefont {Ziegler}, \citenamefont {Golubev}, \citenamefont {Antonov}, \citenamefont {Il’ichev},\ and\ \citenamefont {Astafiev}}]{shaikhaidarovCurrentSteps2022}%
  \BibitemOpen
  \bibfield  {author} {\bibinfo {author} {\bibfnamefont {R.~S.}\ \bibnamefont {Shaikhaidarov}}, \bibinfo {author} {\bibfnamefont {K.~H.}\ \bibnamefont {Kim}}, \bibinfo {author} {\bibfnamefont {J.~W.}\ \bibnamefont {Dunstan}}, \bibinfo {author} {\bibfnamefont {I.~V.}\ \bibnamefont {Antonov}}, \bibinfo {author} {\bibfnamefont {S.}~\bibnamefont {Linzen}}, \bibinfo {author} {\bibfnamefont {M.}~\bibnamefont {Ziegler}}, \bibinfo {author} {\bibfnamefont {D.~S.}\ \bibnamefont {Golubev}}, \bibinfo {author} {\bibfnamefont {V.~N.}\ \bibnamefont {Antonov}}, \bibinfo {author} {\bibfnamefont {E.~V.}\ \bibnamefont {Il’ichev}},\ and\ \bibinfo {author} {\bibfnamefont {O.~V.}\ \bibnamefont {Astafiev}},\ }\bibfield  {title} {\bibinfo {title} {Quantized current steps due to the a.c. coherent quantum phase-slip effect},\ }\href {https://doi.org/10.1038/s41586-022-04947-z} {\bibfield  {journal} {\bibinfo  {journal} {Nature}\ }\textbf {\bibinfo {volume} {608}},\ \bibinfo {pages} {45} (\bibinfo {year} {2022})}\BibitemShut
  {NoStop}%
\bibitem [{\citenamefont {Crescini}\ \emph {et~al.}(2023)\citenamefont {Crescini}, \citenamefont {Cailleaux}, \citenamefont {Guichard}, \citenamefont {Naud}, \citenamefont {Buisson}, \citenamefont {W.~Murch},\ and\ \citenamefont {Roch}}]{cresciniDual2023}%
  \BibitemOpen
  \bibfield  {author} {\bibinfo {author} {\bibfnamefont {N.}~\bibnamefont {Crescini}}, \bibinfo {author} {\bibfnamefont {S.}~\bibnamefont {Cailleaux}}, \bibinfo {author} {\bibfnamefont {W.}~\bibnamefont {Guichard}}, \bibinfo {author} {\bibfnamefont {C.}~\bibnamefont {Naud}}, \bibinfo {author} {\bibfnamefont {O.}~\bibnamefont {Buisson}}, \bibinfo {author} {\bibfnamefont {K.}~\bibnamefont {W.~Murch}},\ and\ \bibinfo {author} {\bibfnamefont {N.}~\bibnamefont {Roch}},\ }\bibfield  {title} {\bibinfo {title} {Evidence of dual shapiro steps in a josephson junction array},\ }\href {https://doi.org/10.1038/s41567-023-01961-4} {\bibfield  {journal} {\bibinfo  {journal} {Nat. Phys.}\ }\textbf {\bibinfo {volume} {19}},\ \bibinfo {pages} {851} (\bibinfo {year} {2023})}\BibitemShut {NoStop}%
\bibitem [{\citenamefont {Kaap}\ \emph {et~al.}(2024)\citenamefont {Kaap}, \citenamefont {Kissling}, \citenamefont {Gaydamachenko}, \citenamefont {Grünhaupt},\ and\ \citenamefont {Lotkhov}}]{kaapShapiro2024}%
  \BibitemOpen
  \bibfield  {author} {\bibinfo {author} {\bibfnamefont {F.}~\bibnamefont {Kaap}}, \bibinfo {author} {\bibfnamefont {C.}~\bibnamefont {Kissling}}, \bibinfo {author} {\bibfnamefont {V.}~\bibnamefont {Gaydamachenko}}, \bibinfo {author} {\bibfnamefont {L.}~\bibnamefont {Grünhaupt}},\ and\ \bibinfo {author} {\bibfnamefont {S.}~\bibnamefont {Lotkhov}},\ }\bibfield  {title} {\bibinfo {title} {Demonstration of dual shapiro steps in small josephson junctions},\ }\href {https://doi.org/10.1038/s41467-024-53011-z} {\bibfield  {journal} {\bibinfo  {journal} {Nat. Commun.}\ }\textbf {\bibinfo {volume} {15}},\ \bibinfo {pages} {8726} (\bibinfo {year} {2024})}\BibitemShut {NoStop}%
\bibitem [{\citenamefont {Pechenezhskiy}\ \emph {et~al.}(2020)\citenamefont {Pechenezhskiy}, \citenamefont {Mencia}, \citenamefont {Nguyen}, \citenamefont {Lin},\ and\ \citenamefont {Manucharyan}}]{pechenezhskiySuperconding2020}%
  \BibitemOpen
  \bibfield  {author} {\bibinfo {author} {\bibfnamefont {I.~V.}\ \bibnamefont {Pechenezhskiy}}, \bibinfo {author} {\bibfnamefont {R.~A.}\ \bibnamefont {Mencia}}, \bibinfo {author} {\bibfnamefont {L.~B.}\ \bibnamefont {Nguyen}}, \bibinfo {author} {\bibfnamefont {Y.-H.}\ \bibnamefont {Lin}},\ and\ \bibinfo {author} {\bibfnamefont {V.~E.}\ \bibnamefont {Manucharyan}},\ }\bibfield  {title} {\bibinfo {title} {The superconducting quasicharge qubit},\ }\href {https://doi.org/10.1038/s41586-020-2687-9} {\bibfield  {journal} {\bibinfo  {journal} {Nature}\ }\textbf {\bibinfo {volume} {585}},\ \bibinfo {pages} {368} (\bibinfo {year} {2020})}\BibitemShut {NoStop}%
\bibitem [{\citenamefont {J{\"u}nger}\ \emph {et~al.}(2025)\citenamefont {J{\"u}nger}, \citenamefont {Chistolini}, \citenamefont {Nguyen}, \citenamefont {Kim}, \citenamefont {Chen}, \citenamefont {Ersevim}, \citenamefont {Livingston}, \citenamefont {Koolstra}, \citenamefont {Santiago},\ and\ \citenamefont {Siddiqi}}]{jungerSuspended2025}%
  \BibitemOpen
  \bibfield  {author} {\bibinfo {author} {\bibfnamefont {C.}~\bibnamefont {J{\"u}nger}}, \bibinfo {author} {\bibfnamefont {T.}~\bibnamefont {Chistolini}}, \bibinfo {author} {\bibfnamefont {L.~B.}\ \bibnamefont {Nguyen}}, \bibinfo {author} {\bibfnamefont {H.}~\bibnamefont {Kim}}, \bibinfo {author} {\bibfnamefont {L.}~\bibnamefont {Chen}}, \bibinfo {author} {\bibfnamefont {T.}~\bibnamefont {Ersevim}}, \bibinfo {author} {\bibfnamefont {W.}~\bibnamefont {Livingston}}, \bibinfo {author} {\bibfnamefont {G.}~\bibnamefont {Koolstra}}, \bibinfo {author} {\bibfnamefont {D.~I.}\ \bibnamefont {Santiago}},\ and\ \bibinfo {author} {\bibfnamefont {I.}~\bibnamefont {Siddiqi}},\ }\bibfield  {title} {\bibinfo {title} {Implementation of scalable suspended superinductors},\ }\href {https://doi.org/10.1063/5.0250341} {\bibfield  {journal} {\bibinfo  {journal} {Appl. Phys. Lett.}\ }\textbf {\bibinfo {volume} {126}},\ \bibinfo {pages} {044003} (\bibinfo {year} {2025})}\BibitemShut {NoStop}%
\bibitem [{\citenamefont {Frasca}\ \emph {et~al.}(2023)\citenamefont {Frasca}, \citenamefont {Arabadzhiev}, \citenamefont {De~Puechredon}, \citenamefont {Oppliger}, \citenamefont {Jouanny}, \citenamefont {Musio}, \citenamefont {Scigliuzzo}, \citenamefont {Minganti}, \citenamefont {Scarlino},\ and\ \citenamefont {Charbon}}]{frascaNbN2023}%
  \BibitemOpen
  \bibfield  {author} {\bibinfo {author} {\bibfnamefont {S.}~\bibnamefont {Frasca}}, \bibinfo {author} {\bibfnamefont {I.}~\bibnamefont {Arabadzhiev}}, \bibinfo {author} {\bibfnamefont {S.~B.}\ \bibnamefont {De~Puechredon}}, \bibinfo {author} {\bibfnamefont {F.}~\bibnamefont {Oppliger}}, \bibinfo {author} {\bibfnamefont {V.}~\bibnamefont {Jouanny}}, \bibinfo {author} {\bibfnamefont {R.}~\bibnamefont {Musio}}, \bibinfo {author} {\bibfnamefont {M.}~\bibnamefont {Scigliuzzo}}, \bibinfo {author} {\bibfnamefont {F.}~\bibnamefont {Minganti}}, \bibinfo {author} {\bibfnamefont {P.}~\bibnamefont {Scarlino}},\ and\ \bibinfo {author} {\bibfnamefont {E.}~\bibnamefont {Charbon}},\ }\bibfield  {title} {\bibinfo {title} {Nb{N} films with high kinetic inductance for high-quality compact superconducting resonators},\ }\href {https://doi.org/10.1103/PhysRevApplied.20.044021} {\bibfield  {journal} {\bibinfo  {journal} {Phys. Rev. Appl.}\ }\textbf {\bibinfo {volume} {20}},\ \bibinfo {pages} {044021} (\bibinfo {year}
  {2023})}\BibitemShut {NoStop}%
\bibitem [{\citenamefont {Peruzzo}\ \emph {et~al.}(2020)\citenamefont {Peruzzo}, \citenamefont {Trioni}, \citenamefont {Hassani}, \citenamefont {Zemlicka},\ and\ \citenamefont {Fink}}]{peruzzoSurpassing2020}%
  \BibitemOpen
  \bibfield  {author} {\bibinfo {author} {\bibfnamefont {M.}~\bibnamefont {Peruzzo}}, \bibinfo {author} {\bibfnamefont {A.}~\bibnamefont {Trioni}}, \bibinfo {author} {\bibfnamefont {F.}~\bibnamefont {Hassani}}, \bibinfo {author} {\bibfnamefont {M.}~\bibnamefont {Zemlicka}},\ and\ \bibinfo {author} {\bibfnamefont {J.~M.}\ \bibnamefont {Fink}},\ }\bibfield  {title} {\bibinfo {title} {Surpassing the resistance quantum with a geometric superinductor},\ }\href {https://doi.org/10.1103/PhysRevApplied.14.044055} {\bibfield  {journal} {\bibinfo  {journal} {Phys. Rev. Appl.}\ }\textbf {\bibinfo {volume} {14}},\ \bibinfo {pages} {044055} (\bibinfo {year} {2020})}\BibitemShut {NoStop}%
\bibitem [{\citenamefont {Medahinne}\ \emph {et~al.}()\citenamefont {Medahinne}, \citenamefont {Kandel}, \citenamefont {Magar}, \citenamefont {Champion}, \citenamefont {Nichol},\ and\ \citenamefont {Blok}}]{medahinneMagnetic2024}%
  \BibitemOpen
  \bibfield  {author} {\bibinfo {author} {\bibfnamefont {M.}~\bibnamefont {Medahinne}}, \bibinfo {author} {\bibfnamefont {Y.~P.}\ \bibnamefont {Kandel}}, \bibinfo {author} {\bibfnamefont {S.~T.}\ \bibnamefont {Magar}}, \bibinfo {author} {\bibfnamefont {E.}~\bibnamefont {Champion}}, \bibinfo {author} {\bibfnamefont {J.~M.}\ \bibnamefont {Nichol}},\ and\ \bibinfo {author} {\bibfnamefont {M.~S.}\ \bibnamefont {Blok}},\ }\href@noop {} {\bibinfo {title} {Magnetic field tolerant superconducting spiral resonators for circuit {QED}}},\ \bibinfo {note} {\href{https://arxiv.org/abs/2406.10386}{arXiv (2024), 2406.10386}}\BibitemShut {NoStop}%
\bibitem [{\citenamefont {{Pita-Vidal}}\ \emph {et~al.}(2024)\citenamefont {{Pita-Vidal}}, \citenamefont {Wesdorp}, \citenamefont {Splitthoff}, \citenamefont {Bargerbos}, \citenamefont {Liu}, \citenamefont {Kouwenhoven},\ and\ \citenamefont {Andersen}}]{pita-vidalAndreev2024}%
  \BibitemOpen
  \bibfield  {author} {\bibinfo {author} {\bibfnamefont {M.}~\bibnamefont {{Pita-Vidal}}}, \bibinfo {author} {\bibfnamefont {J.~J.}\ \bibnamefont {Wesdorp}}, \bibinfo {author} {\bibfnamefont {L.~J.}\ \bibnamefont {Splitthoff}}, \bibinfo {author} {\bibfnamefont {A.}~\bibnamefont {Bargerbos}}, \bibinfo {author} {\bibfnamefont {Y.}~\bibnamefont {Liu}}, \bibinfo {author} {\bibfnamefont {L.~P.}\ \bibnamefont {Kouwenhoven}},\ and\ \bibinfo {author} {\bibfnamefont {C.~K.}\ \bibnamefont {Andersen}},\ }\bibfield  {title} {\bibinfo {title} {Strong tunable coupling between two distant superconducting spin qubits},\ }\href {https://doi.org/10.1038/s41567-024-02497-x} {\bibfield  {journal} {\bibinfo  {journal} {Nat. Phys.}\ }\textbf {\bibinfo {volume} {20}},\ \bibinfo {pages} {1158} (\bibinfo {year} {2024})}\BibitemShut {NoStop}%
\bibitem [{\citenamefont {Larsen}\ \emph {et~al.}(2015)\citenamefont {Larsen}, \citenamefont {Petersson}, \citenamefont {Kuemmeth}, \citenamefont {Jespersen}, \citenamefont {Krogstrup}, \citenamefont {Nyg{\aa}rd},\ and\ \citenamefont {Marcus}}]{larsenNanowireBQubit2015}%
  \BibitemOpen
  \bibfield  {author} {\bibinfo {author} {\bibfnamefont {T.~W.}\ \bibnamefont {Larsen}}, \bibinfo {author} {\bibfnamefont {K.~D.}\ \bibnamefont {Petersson}}, \bibinfo {author} {\bibfnamefont {F.}~\bibnamefont {Kuemmeth}}, \bibinfo {author} {\bibfnamefont {T.~S.}\ \bibnamefont {Jespersen}}, \bibinfo {author} {\bibfnamefont {P.}~\bibnamefont {Krogstrup}}, \bibinfo {author} {\bibfnamefont {J.}~\bibnamefont {Nyg{\aa}rd}},\ and\ \bibinfo {author} {\bibfnamefont {C.~M.}\ \bibnamefont {Marcus}},\ }\bibfield  {title} {\bibinfo {title} {Semiconductor-nanowire-based superconducting qubit},\ }\href {https://doi.org/10.1103/PhysRevLett.115.127001} {\bibfield  {journal} {\bibinfo  {journal} {Phys. Rev. Lett.}\ }\textbf {\bibinfo {volume} {115}},\ \bibinfo {pages} {127001} (\bibinfo {year} {2015})}\BibitemShut {NoStop}%
\bibitem [{\citenamefont {Valentini}\ \emph {et~al.}(2021)\citenamefont {Valentini}, \citenamefont {Pe{\~n}aranda}, \citenamefont {Hofmann}, \citenamefont {Brauns}, \citenamefont {Hauschild}, \citenamefont {Krogstrup}, \citenamefont {{San-Jose}}, \citenamefont {Prada}, \citenamefont {Aguado},\ and\ \citenamefont {Katsaros}}]{valentiniNanowireQubit2021}%
  \BibitemOpen
  \bibfield  {author} {\bibinfo {author} {\bibfnamefont {M.}~\bibnamefont {Valentini}}, \bibinfo {author} {\bibfnamefont {F.}~\bibnamefont {Pe{\~n}aranda}}, \bibinfo {author} {\bibfnamefont {A.}~\bibnamefont {Hofmann}}, \bibinfo {author} {\bibfnamefont {M.}~\bibnamefont {Brauns}}, \bibinfo {author} {\bibfnamefont {R.}~\bibnamefont {Hauschild}}, \bibinfo {author} {\bibfnamefont {P.}~\bibnamefont {Krogstrup}}, \bibinfo {author} {\bibfnamefont {P.}~\bibnamefont {{San-Jose}}}, \bibinfo {author} {\bibfnamefont {E.}~\bibnamefont {Prada}}, \bibinfo {author} {\bibfnamefont {R.}~\bibnamefont {Aguado}},\ and\ \bibinfo {author} {\bibfnamefont {G.}~\bibnamefont {Katsaros}},\ }\bibfield  {title} {\bibinfo {title} {Nontopological zero-bias peaks in full-shell nanowires induced by flux-tunable {{Andreev}} states},\ }\href {https://doi.org/10.1126/science.abf1513} {\bibfield  {journal} {\bibinfo  {journal} {Science}\ }\textbf {\bibinfo {volume} {373}},\ \bibinfo {pages} {82} (\bibinfo {year} {2021})}\BibitemShut {NoStop}%
\bibitem [{\citenamefont {Luthi}\ \emph {et~al.}(2018)\citenamefont {Luthi}, \citenamefont {Stavenga}, \citenamefont {Enzing}, \citenamefont {Bruno}, \citenamefont {Dickel}, \citenamefont {Langford}, \citenamefont {Rol}, \citenamefont {Jespersen}, \citenamefont {Nyg{\aa}rd}, \citenamefont {Krogstrup},\ and\ \citenamefont {DiCarlo}}]{luthiNanowireTransmon2018}%
  \BibitemOpen
  \bibfield  {author} {\bibinfo {author} {\bibfnamefont {F.}~\bibnamefont {Luthi}}, \bibinfo {author} {\bibfnamefont {T.}~\bibnamefont {Stavenga}}, \bibinfo {author} {\bibfnamefont {O.~W.}\ \bibnamefont {Enzing}}, \bibinfo {author} {\bibfnamefont {A.}~\bibnamefont {Bruno}}, \bibinfo {author} {\bibfnamefont {C.}~\bibnamefont {Dickel}}, \bibinfo {author} {\bibfnamefont {N.~K.}\ \bibnamefont {Langford}}, \bibinfo {author} {\bibfnamefont {M.~A.}\ \bibnamefont {Rol}}, \bibinfo {author} {\bibfnamefont {T.~S.}\ \bibnamefont {Jespersen}}, \bibinfo {author} {\bibfnamefont {J.}~\bibnamefont {Nyg{\aa}rd}}, \bibinfo {author} {\bibfnamefont {P.}~\bibnamefont {Krogstrup}},\ and\ \bibinfo {author} {\bibfnamefont {L.}~\bibnamefont {DiCarlo}},\ }\bibfield  {title} {\bibinfo {title} {Evolution of nanowire transmon qubits and their coherence in a magnetic field},\ }\href {https://doi.org/10.1103/PhysRevLett.120.100502} {\bibfield  {journal} {\bibinfo  {journal} {Phys. Rev. Lett.}\ }\textbf {\bibinfo {volume} {120}},\ \bibinfo
  {pages} {100502} (\bibinfo {year} {2018})}\BibitemShut {NoStop}%
\bibitem [{\citenamefont {Niepce}\ \emph {et~al.}(2019)\citenamefont {Niepce}, \citenamefont {Burnett},\ and\ \citenamefont {Bylander}}]{niepceHigh2019}%
  \BibitemOpen
  \bibfield  {author} {\bibinfo {author} {\bibfnamefont {D.}~\bibnamefont {Niepce}}, \bibinfo {author} {\bibfnamefont {J.}~\bibnamefont {Burnett}},\ and\ \bibinfo {author} {\bibfnamefont {J.}~\bibnamefont {Bylander}},\ }\bibfield  {title} {\bibinfo {title} {High kinetic inductance {N}b{N} nanowire superinductors},\ }\href {https://doi.org/10.1103/PhysRevApplied.11.044014} {\bibfield  {journal} {\bibinfo  {journal} {Phys. Rev. Appl.}\ }\textbf {\bibinfo {volume} {11}},\ \bibinfo {pages} {044014} (\bibinfo {year} {2019})}\BibitemShut {NoStop}%
\bibitem [{\citenamefont {Yang}\ \emph {et~al.}(2024)\citenamefont {Yang}, \citenamefont {He}, \citenamefont {Gao}, \citenamefont {Chen}, \citenamefont {Wu}, \citenamefont {Wang}, \citenamefont {Mu}, \citenamefont {Peng},\ and\ \citenamefont {Lin}}]{yangNbTiNResonator2024}%
  \BibitemOpen
  \bibfield  {author} {\bibinfo {author} {\bibfnamefont {M.}~\bibnamefont {Yang}}, \bibinfo {author} {\bibfnamefont {X.}~\bibnamefont {He}}, \bibinfo {author} {\bibfnamefont {W.}~\bibnamefont {Gao}}, \bibinfo {author} {\bibfnamefont {J.}~\bibnamefont {Chen}}, \bibinfo {author} {\bibfnamefont {Y.}~\bibnamefont {Wu}}, \bibinfo {author} {\bibfnamefont {X.}~\bibnamefont {Wang}}, \bibinfo {author} {\bibfnamefont {G.}~\bibnamefont {Mu}}, \bibinfo {author} {\bibfnamefont {W.}~\bibnamefont {Peng}},\ and\ \bibinfo {author} {\bibfnamefont {Z.}~\bibnamefont {Lin}},\ }\bibfield  {title} {\bibinfo {title} {Kinetic inductance compact resonator with nbtin micronwires},\ }\bibfield  {journal} {\bibinfo  {journal} {AIP Adv.}\ }\textbf {\bibinfo {volume} {14}},\ \href {https://doi.org/10.1063/5.0220296} {10.1063/5.0220296} (\bibinfo {year} {2024})\BibitemShut {NoStop}%
\bibitem [{\citenamefont {Shearrow}\ \emph {et~al.}(2018)\citenamefont {Shearrow}, \citenamefont {Koolstra}, \citenamefont {Whiteley}, \citenamefont {Earnest}, \citenamefont {Barry}, \citenamefont {Heremans}, \citenamefont {Awschalom}, \citenamefont {Shirokoff},\ and\ \citenamefont {Schuster}}]{shearrowAtomic2018}%
  \BibitemOpen
  \bibfield  {author} {\bibinfo {author} {\bibfnamefont {A.}~\bibnamefont {Shearrow}}, \bibinfo {author} {\bibfnamefont {G.}~\bibnamefont {Koolstra}}, \bibinfo {author} {\bibfnamefont {S.~J.}\ \bibnamefont {Whiteley}}, \bibinfo {author} {\bibfnamefont {N.}~\bibnamefont {Earnest}}, \bibinfo {author} {\bibfnamefont {P.~S.}\ \bibnamefont {Barry}}, \bibinfo {author} {\bibfnamefont {F.~J.}\ \bibnamefont {Heremans}}, \bibinfo {author} {\bibfnamefont {D.~D.}\ \bibnamefont {Awschalom}}, \bibinfo {author} {\bibfnamefont {E.}~\bibnamefont {Shirokoff}},\ and\ \bibinfo {author} {\bibfnamefont {D.~I.}\ \bibnamefont {Schuster}},\ }\bibfield  {title} {\bibinfo {title} {Atomic layer deposition of titanium nitride for quantum circuits},\ }\href {https://doi.org/10.1063/1.5053461} {\bibfield  {journal} {\bibinfo  {journal} {Appl. Phys. Lett.}\ }\textbf {\bibinfo {volume} {113}},\ \bibinfo {pages} {212601} (\bibinfo {year} {2018})}\BibitemShut {NoStop}%
\bibitem [{\citenamefont {Amin}\ \emph {et~al.}(2022)\citenamefont {Amin}, \citenamefont {Ladner}, \citenamefont {Jourdan}, \citenamefont {Hentz}, \citenamefont {Roch},\ and\ \citenamefont {Renard}}]{aminLossMechanisms2022}%
  \BibitemOpen
  \bibfield  {author} {\bibinfo {author} {\bibfnamefont {K.~R.}\ \bibnamefont {Amin}}, \bibinfo {author} {\bibfnamefont {C.}~\bibnamefont {Ladner}}, \bibinfo {author} {\bibfnamefont {G.}~\bibnamefont {Jourdan}}, \bibinfo {author} {\bibfnamefont {S.}~\bibnamefont {Hentz}}, \bibinfo {author} {\bibfnamefont {N.}~\bibnamefont {Roch}},\ and\ \bibinfo {author} {\bibfnamefont {J.}~\bibnamefont {Renard}},\ }\bibfield  {title} {\bibinfo {title} {Loss mechanisms in tin high impedance superconducting microwave circuits},\ }\href {https://doi.org/10.1063/5.0086019} {\bibfield  {journal} {\bibinfo  {journal} {Appl. Phys. Lett.}\ }\textbf {\bibinfo {volume} {120}},\ \bibinfo {pages} {164001} (\bibinfo {year} {2022})}\BibitemShut {NoStop}%
\bibitem [{\citenamefont {Vissers}\ \emph {et~al.}(2010)\citenamefont {Vissers}, \citenamefont {Gao}, \citenamefont {Wisbey}, \citenamefont {Hite}, \citenamefont {Tsuei}, \citenamefont {Corcoles}, \citenamefont {Steffen},\ and\ \citenamefont {Pappas}}]{vissersLowLossTiN2010}%
  \BibitemOpen
  \bibfield  {author} {\bibinfo {author} {\bibfnamefont {M.~R.}\ \bibnamefont {Vissers}}, \bibinfo {author} {\bibfnamefont {J.}~\bibnamefont {Gao}}, \bibinfo {author} {\bibfnamefont {D.~S.}\ \bibnamefont {Wisbey}}, \bibinfo {author} {\bibfnamefont {D.~A.}\ \bibnamefont {Hite}}, \bibinfo {author} {\bibfnamefont {C.~C.}\ \bibnamefont {Tsuei}}, \bibinfo {author} {\bibfnamefont {A.~D.}\ \bibnamefont {Corcoles}}, \bibinfo {author} {\bibfnamefont {M.}~\bibnamefont {Steffen}},\ and\ \bibinfo {author} {\bibfnamefont {D.~P.}\ \bibnamefont {Pappas}},\ }\bibfield  {title} {\bibinfo {title} {Low loss superconducting titanium nitride coplanar waveguide resonators},\ }\href {https://doi.org/10.1063/1.3517252} {\bibfield  {journal} {\bibinfo  {journal} {Appl. Phys. Lett.}\ }\textbf {\bibinfo {volume} {97}},\ \bibinfo {pages} {232509} (\bibinfo {year} {2010})}\BibitemShut {NoStop}%
\bibitem [{\citenamefont {Samkharadze}\ \emph {et~al.}(2016)\citenamefont {Samkharadze}, \citenamefont {Bruno}, \citenamefont {Scarlino}, \citenamefont {Zheng}, \citenamefont {DiVincenzo}, \citenamefont {DiCarlo},\ and\ \citenamefont {Vandersypen}}]{samkharadzeHigh2016}%
  \BibitemOpen
  \bibfield  {author} {\bibinfo {author} {\bibfnamefont {N.}~\bibnamefont {Samkharadze}}, \bibinfo {author} {\bibfnamefont {A.}~\bibnamefont {Bruno}}, \bibinfo {author} {\bibfnamefont {P.}~\bibnamefont {Scarlino}}, \bibinfo {author} {\bibfnamefont {G.}~\bibnamefont {Zheng}}, \bibinfo {author} {\bibfnamefont {D.~P.}\ \bibnamefont {DiVincenzo}}, \bibinfo {author} {\bibfnamefont {L.}~\bibnamefont {DiCarlo}},\ and\ \bibinfo {author} {\bibfnamefont {L.~M.~K.}\ \bibnamefont {Vandersypen}},\ }\bibfield  {title} {\bibinfo {title} {High-kinetic-inductance superconducting nanowire resonators for circuit {QED} in a magnetic field},\ }\href {https://doi.org/10.1103/PhysRevApplied.5.044004} {\bibfield  {journal} {\bibinfo  {journal} {Phys. Rev. Appl.}\ }\textbf {\bibinfo {volume} {5}},\ \bibinfo {pages} {044004} (\bibinfo {year} {2016})}\BibitemShut {NoStop}%
\bibitem [{\citenamefont {Kroll}\ \emph {et~al.}(2019)\citenamefont {Kroll}, \citenamefont {Borsoi}, \citenamefont {Van Der~Enden}, \citenamefont {Uilhoorn}, \citenamefont {De~Jong}, \citenamefont {Quintero-Pérez}, \citenamefont {Van~Woerkom}, \citenamefont {Bruno}, \citenamefont {Plissard}, \citenamefont {Car}, \citenamefont {Bakkers}, \citenamefont {Cassidy},\ and\ \citenamefont {Kouwenhoven}}]{krollNbTiN2019}%
  \BibitemOpen
  \bibfield  {author} {\bibinfo {author} {\bibfnamefont {J.~G.}\ \bibnamefont {Kroll}}, \bibinfo {author} {\bibfnamefont {F.}~\bibnamefont {Borsoi}}, \bibinfo {author} {\bibfnamefont {K.~L.}\ \bibnamefont {Van Der~Enden}}, \bibinfo {author} {\bibfnamefont {W.}~\bibnamefont {Uilhoorn}}, \bibinfo {author} {\bibfnamefont {D.}~\bibnamefont {De~Jong}}, \bibinfo {author} {\bibfnamefont {M.}~\bibnamefont {Quintero-Pérez}}, \bibinfo {author} {\bibfnamefont {D.~J.}\ \bibnamefont {Van~Woerkom}}, \bibinfo {author} {\bibfnamefont {A.}~\bibnamefont {Bruno}}, \bibinfo {author} {\bibfnamefont {S.~R.}\ \bibnamefont {Plissard}}, \bibinfo {author} {\bibfnamefont {D.}~\bibnamefont {Car}}, \bibinfo {author} {\bibfnamefont {E.~P. A.~M.}\ \bibnamefont {Bakkers}}, \bibinfo {author} {\bibfnamefont {M.~C.}\ \bibnamefont {Cassidy}},\ and\ \bibinfo {author} {\bibfnamefont {L.~P.}\ \bibnamefont {Kouwenhoven}},\ }\bibfield  {title} {\bibinfo {title} {Magnetic-field-resilient superconducting coplanar-waveguide resonators for hybrid
  circuit quantum electrodynamics experiments},\ }\href {https://doi.org/10.1103/PhysRevApplied.11.064053} {\bibfield  {journal} {\bibinfo  {journal} {Phys. Rev. Applied}\ }\textbf {\bibinfo {volume} {11}},\ \bibinfo {pages} {064053} (\bibinfo {year} {2019})}\BibitemShut {NoStop}%
\bibitem [{\citenamefont {Müller}\ \emph {et~al.}(2022)\citenamefont {Müller}, \citenamefont {Luschmann}, \citenamefont {Faltermeier}, \citenamefont {Weichselbaumer}, \citenamefont {Koch}, \citenamefont {Huber}, \citenamefont {Schumacher}, \citenamefont {Ubbelohde}, \citenamefont {Reifert}, \citenamefont {Scheller}, \citenamefont {Deppe}, \citenamefont {Marx}, \citenamefont {Filipp}, \citenamefont {Althammer}, \citenamefont {Gross},\ and\ \citenamefont {Huebl}}]{mullerNbTiNquality2022}%
  \BibitemOpen
  \bibfield  {author} {\bibinfo {author} {\bibfnamefont {M.}~\bibnamefont {Müller}}, \bibinfo {author} {\bibfnamefont {T.}~\bibnamefont {Luschmann}}, \bibinfo {author} {\bibfnamefont {A.}~\bibnamefont {Faltermeier}}, \bibinfo {author} {\bibfnamefont {S.}~\bibnamefont {Weichselbaumer}}, \bibinfo {author} {\bibfnamefont {L.}~\bibnamefont {Koch}}, \bibinfo {author} {\bibfnamefont {G.~B.~P.}\ \bibnamefont {Huber}}, \bibinfo {author} {\bibfnamefont {H.~W.}\ \bibnamefont {Schumacher}}, \bibinfo {author} {\bibfnamefont {N.}~\bibnamefont {Ubbelohde}}, \bibinfo {author} {\bibfnamefont {D.}~\bibnamefont {Reifert}}, \bibinfo {author} {\bibfnamefont {T.}~\bibnamefont {Scheller}}, \bibinfo {author} {\bibfnamefont {F.}~\bibnamefont {Deppe}}, \bibinfo {author} {\bibfnamefont {A.}~\bibnamefont {Marx}}, \bibinfo {author} {\bibfnamefont {S.}~\bibnamefont {Filipp}}, \bibinfo {author} {\bibfnamefont {M.}~\bibnamefont {Althammer}}, \bibinfo {author} {\bibfnamefont {R.}~\bibnamefont {Gross}},\ and\ \bibinfo {author}
  {\bibfnamefont {H.}~\bibnamefont {Huebl}},\ }\bibfield  {title} {\bibinfo {title} {Magnetic field robust high quality factor nbtin superconducting microwave resonators},\ }\href {https://doi.org/10.1088/2633-4356/ac50f8} {\bibfield  {journal} {\bibinfo  {journal} {Mater. Quantum. Technol.}\ }\textbf {\bibinfo {volume} {2}},\ \bibinfo {pages} {015002} (\bibinfo {year} {2022})}\BibitemShut {NoStop}%
\bibitem [{\citenamefont {Gao}\ \emph {et~al.}(2022)\citenamefont {Gao}, \citenamefont {Ku}, \citenamefont {Deng}, \citenamefont {Yu}, \citenamefont {Xia}, \citenamefont {Wu}, \citenamefont {Song}, \citenamefont {Wang}, \citenamefont {Miao}, \citenamefont {Zhang}, \citenamefont {Lin}, \citenamefont {Shi}, \citenamefont {Zhao},\ and\ \citenamefont {Deng}}]{gaoNbAlN2022}%
  \BibitemOpen
  \bibfield  {author} {\bibinfo {author} {\bibfnamefont {R.}~\bibnamefont {Gao}}, \bibinfo {author} {\bibfnamefont {H.}~\bibnamefont {Ku}}, \bibinfo {author} {\bibfnamefont {H.}~\bibnamefont {Deng}}, \bibinfo {author} {\bibfnamefont {W.}~\bibnamefont {Yu}}, \bibinfo {author} {\bibfnamefont {T.}~\bibnamefont {Xia}}, \bibinfo {author} {\bibfnamefont {F.}~\bibnamefont {Wu}}, \bibinfo {author} {\bibfnamefont {Z.}~\bibnamefont {Song}}, \bibinfo {author} {\bibfnamefont {M.}~\bibnamefont {Wang}}, \bibinfo {author} {\bibfnamefont {X.}~\bibnamefont {Miao}}, \bibinfo {author} {\bibfnamefont {C.}~\bibnamefont {Zhang}}, \bibinfo {author} {\bibfnamefont {Y.}~\bibnamefont {Lin}}, \bibinfo {author} {\bibfnamefont {Y.}~\bibnamefont {Shi}}, \bibinfo {author} {\bibfnamefont {H.}~\bibnamefont {Zhao}},\ and\ \bibinfo {author} {\bibfnamefont {C.}~\bibnamefont {Deng}},\ }\bibfield  {title} {\bibinfo {title} {Ultrahigh kinetic inductance superconducting materials from spinodal decomposition},\ }\href
  {https://doi.org/10.1002/adma.202201268} {\bibfield  {journal} {\bibinfo  {journal} {Adv. Mater.}\ }\textbf {\bibinfo {volume} {34}},\ \bibinfo {pages} {2201268} (\bibinfo {year} {2022})}\BibitemShut {NoStop}%
\bibitem [{\citenamefont {Rotzinger}\ \emph {et~al.}(2016)\citenamefont {Rotzinger}, \citenamefont {Skacel}, \citenamefont {Pfirrmann}, \citenamefont {Voss}, \citenamefont {Münzberg}, \citenamefont {Probst}, \citenamefont {Bushev}, \citenamefont {Weides}, \citenamefont {Ustinov},\ and\ \citenamefont {Mooij}}]{rotzingergral2017}%
  \BibitemOpen
  \bibfield  {author} {\bibinfo {author} {\bibfnamefont {H.}~\bibnamefont {Rotzinger}}, \bibinfo {author} {\bibfnamefont {S.~T.}\ \bibnamefont {Skacel}}, \bibinfo {author} {\bibfnamefont {M.}~\bibnamefont {Pfirrmann}}, \bibinfo {author} {\bibfnamefont {J.~N.}\ \bibnamefont {Voss}}, \bibinfo {author} {\bibfnamefont {J.}~\bibnamefont {Münzberg}}, \bibinfo {author} {\bibfnamefont {S.}~\bibnamefont {Probst}}, \bibinfo {author} {\bibfnamefont {P.}~\bibnamefont {Bushev}}, \bibinfo {author} {\bibfnamefont {M.~P.}\ \bibnamefont {Weides}}, \bibinfo {author} {\bibfnamefont {A.~V.}\ \bibnamefont {Ustinov}},\ and\ \bibinfo {author} {\bibfnamefont {J.~E.}\ \bibnamefont {Mooij}},\ }\bibfield  {title} {\bibinfo {title} {Aluminium-oxide wires for superconducting high kinetic inductance circuits},\ }\href {https://doi.org/10.1088/0953-2048/30/2/025002} {\bibfield  {journal} {\bibinfo  {journal} {Supercond. Sci. Technol.}\ }\textbf {\bibinfo {volume} {30}},\ \bibinfo {pages} {025002} (\bibinfo {year} {2016})}\BibitemShut
  {NoStop}%
\bibitem [{\citenamefont {Kamenov}\ \emph {et~al.}(2020)\citenamefont {Kamenov}, \citenamefont {Lu}, \citenamefont {Kalashnikov}, \citenamefont {DiNapoli}, \citenamefont {Bell},\ and\ \citenamefont {Gershenson}}]{kamenovMeandered2020}%
  \BibitemOpen
  \bibfield  {author} {\bibinfo {author} {\bibfnamefont {P.}~\bibnamefont {Kamenov}}, \bibinfo {author} {\bibfnamefont {W.-S.}\ \bibnamefont {Lu}}, \bibinfo {author} {\bibfnamefont {K.}~\bibnamefont {Kalashnikov}}, \bibinfo {author} {\bibfnamefont {T.}~\bibnamefont {DiNapoli}}, \bibinfo {author} {\bibfnamefont {M.~T.}\ \bibnamefont {Bell}},\ and\ \bibinfo {author} {\bibfnamefont {M.~E.}\ \bibnamefont {Gershenson}},\ }\bibfield  {title} {\bibinfo {title} {Granular aluminum meandered superinductors for quantum circuits},\ }\href {https://doi.org/10.1103/PhysRevApplied.13.054051} {\bibfield  {journal} {\bibinfo  {journal} {Phys. Rev. Appl.}\ }\textbf {\bibinfo {volume} {13}},\ \bibinfo {pages} {054051} (\bibinfo {year} {2020})}\BibitemShut {NoStop}%
\bibitem [{\citenamefont {Janík}\ \emph {et~al.}()\citenamefont {Janík}, \citenamefont {Roux}, \citenamefont {Espinosa}, \citenamefont {Sagi}, \citenamefont {Baghdadi}, \citenamefont {Adletzberger}, \citenamefont {Calcaterra}, \citenamefont {Botifoll}, \citenamefont {Manjón}, \citenamefont {Arbiol}, \citenamefont {Chrastina}, \citenamefont {Isella}, \citenamefont {Pop},\ and\ \citenamefont {Katsaros}}]{janikStrong2024}%
  \BibitemOpen
  \bibfield  {author} {\bibinfo {author} {\bibfnamefont {M.}~\bibnamefont {Janík}}, \bibinfo {author} {\bibfnamefont {K.}~\bibnamefont {Roux}}, \bibinfo {author} {\bibfnamefont {C.~B.}\ \bibnamefont {Espinosa}}, \bibinfo {author} {\bibfnamefont {O.}~\bibnamefont {Sagi}}, \bibinfo {author} {\bibfnamefont {A.}~\bibnamefont {Baghdadi}}, \bibinfo {author} {\bibfnamefont {T.}~\bibnamefont {Adletzberger}}, \bibinfo {author} {\bibfnamefont {S.}~\bibnamefont {Calcaterra}}, \bibinfo {author} {\bibfnamefont {M.}~\bibnamefont {Botifoll}}, \bibinfo {author} {\bibfnamefont {A.~G.}\ \bibnamefont {Manjón}}, \bibinfo {author} {\bibfnamefont {J.}~\bibnamefont {Arbiol}}, \bibinfo {author} {\bibfnamefont {D.}~\bibnamefont {Chrastina}}, \bibinfo {author} {\bibfnamefont {G.}~\bibnamefont {Isella}}, \bibinfo {author} {\bibfnamefont {I.~M.}\ \bibnamefont {Pop}},\ and\ \bibinfo {author} {\bibfnamefont {G.}~\bibnamefont {Katsaros}},\ }\href@noop {} {\bibinfo {title} {Strong charge-photon coupling in planar germanium enabled by
  granular aluminium superinductors}},\ \bibinfo {note} {\href{https://arxiv.org/abs/2407.03079}{arXiv (2024), 2407.03079}}\BibitemShut {NoStop}%
\bibitem [{\citenamefont {Gupta}\ \emph {et~al.}()\citenamefont {Gupta}, \citenamefont {Winkel}, \citenamefont {Thakur}, \citenamefont {van Vlaanderen}, \citenamefont {Wang}, \citenamefont {Ganjam}, \citenamefont {Frunzio},\ and\ \citenamefont {Schoelkopf}}]{guptaLowLoss2024}%
  \BibitemOpen
  \bibfield  {author} {\bibinfo {author} {\bibfnamefont {V.}~\bibnamefont {Gupta}}, \bibinfo {author} {\bibfnamefont {P.}~\bibnamefont {Winkel}}, \bibinfo {author} {\bibfnamefont {N.}~\bibnamefont {Thakur}}, \bibinfo {author} {\bibfnamefont {P.}~\bibnamefont {van Vlaanderen}}, \bibinfo {author} {\bibfnamefont {Y.}~\bibnamefont {Wang}}, \bibinfo {author} {\bibfnamefont {S.}~\bibnamefont {Ganjam}}, \bibinfo {author} {\bibfnamefont {L.}~\bibnamefont {Frunzio}},\ and\ \bibinfo {author} {\bibfnamefont {R.~J.}\ \bibnamefont {Schoelkopf}},\ }\href@noop {} {\bibinfo {title} {Low loss lumped-element inductors made from granular aluminum}},\ \bibinfo {note} {\href{https://arxiv.org/abs/2411.12611}{arXiv (2024), 2411.12611}}\BibitemShut {NoStop}%
\bibitem [{\citenamefont {Xu}\ \emph {et~al.}(2023)\citenamefont {Xu}, \citenamefont {Cheng}, \citenamefont {Wu}, \citenamefont {Liu},\ and\ \citenamefont {Tang}}]{xukerrNbN2023}%
  \BibitemOpen
  \bibfield  {author} {\bibinfo {author} {\bibfnamefont {M.}~\bibnamefont {Xu}}, \bibinfo {author} {\bibfnamefont {R.}~\bibnamefont {Cheng}}, \bibinfo {author} {\bibfnamefont {Y.}~\bibnamefont {Wu}}, \bibinfo {author} {\bibfnamefont {G.}~\bibnamefont {Liu}},\ and\ \bibinfo {author} {\bibfnamefont {H.~X.}\ \bibnamefont {Tang}},\ }\bibfield  {title} {\bibinfo {title} {Magnetic {{Field-Resilient Quantum-Limited Parametric Amplifier}}},\ }\href {https://doi.org/10.1103/PRXQuantum.4.010322} {\bibfield  {journal} {\bibinfo  {journal} {PRX Quantum}\ }\textbf {\bibinfo {volume} {4}},\ \bibinfo {pages} {010322} (\bibinfo {year} {2023})}\BibitemShut {NoStop}%
\bibitem [{\citenamefont {Joshi}\ \emph {et~al.}(2022)\citenamefont {Joshi}, \citenamefont {Chen}, \citenamefont {LeDuc}, \citenamefont {Day},\ and\ \citenamefont {Mirhosseini}}]{joshiKerrNonlinearity2022}%
  \BibitemOpen
  \bibfield  {author} {\bibinfo {author} {\bibfnamefont {C.}~\bibnamefont {Joshi}}, \bibinfo {author} {\bibfnamefont {W.}~\bibnamefont {Chen}}, \bibinfo {author} {\bibfnamefont {H.~G.}\ \bibnamefont {LeDuc}}, \bibinfo {author} {\bibfnamefont {P.~K.}\ \bibnamefont {Day}},\ and\ \bibinfo {author} {\bibfnamefont {M.}~\bibnamefont {Mirhosseini}},\ }\bibfield  {title} {\bibinfo {title} {Strong kinetic-inductance kerr nonlinearity with titanium nitride nanowires},\ }\href {https://doi.org/10.1103/PhysRevApplied.18.064088} {\bibfield  {journal} {\bibinfo  {journal} {Phys. Rev. Applied}\ }\textbf {\bibinfo {volume} {18}},\ \bibinfo {pages} {064088} (\bibinfo {year} {2022})}\BibitemShut {NoStop}%
\bibitem [{\citenamefont {Maleeva}\ \emph {et~al.}(2018)\citenamefont {Maleeva}, \citenamefont {Grünhaupt}, \citenamefont {Klein}, \citenamefont {Levy-Bertrand}, \citenamefont {Dupre}, \citenamefont {Calvo}, \citenamefont {Valenti}, \citenamefont {Winkel}, \citenamefont {Friedrich}, \citenamefont {Wernsdorfer}, \citenamefont {Ustinov}, \citenamefont {Rotzinger}, \citenamefont {Monfardini}, \citenamefont {Fistul},\ and\ \citenamefont {Pop}}]{maleevaCircuit2018}%
  \BibitemOpen
  \bibfield  {author} {\bibinfo {author} {\bibfnamefont {N.}~\bibnamefont {Maleeva}}, \bibinfo {author} {\bibfnamefont {L.}~\bibnamefont {Grünhaupt}}, \bibinfo {author} {\bibfnamefont {T.}~\bibnamefont {Klein}}, \bibinfo {author} {\bibfnamefont {F.}~\bibnamefont {Levy-Bertrand}}, \bibinfo {author} {\bibfnamefont {O.}~\bibnamefont {Dupre}}, \bibinfo {author} {\bibfnamefont {M.}~\bibnamefont {Calvo}}, \bibinfo {author} {\bibfnamefont {F.}~\bibnamefont {Valenti}}, \bibinfo {author} {\bibfnamefont {P.}~\bibnamefont {Winkel}}, \bibinfo {author} {\bibfnamefont {F.}~\bibnamefont {Friedrich}}, \bibinfo {author} {\bibfnamefont {W.}~\bibnamefont {Wernsdorfer}}, \bibinfo {author} {\bibfnamefont {A.~V.}\ \bibnamefont {Ustinov}}, \bibinfo {author} {\bibfnamefont {H.}~\bibnamefont {Rotzinger}}, \bibinfo {author} {\bibfnamefont {A.}~\bibnamefont {Monfardini}}, \bibinfo {author} {\bibfnamefont {M.~V.}\ \bibnamefont {Fistul}},\ and\ \bibinfo {author} {\bibfnamefont {I.~M.}\ \bibnamefont {Pop}},\ }\bibfield  {title} {\bibinfo
  {title} {Circuit quantum electrodynamics of granular aluminum resonators},\ }\href {https://doi.org/10.1038/s41467-018-06386-9} {\bibfield  {journal} {\bibinfo  {journal} {Nat. Commun.}\ }\textbf {\bibinfo {volume} {9}},\ \bibinfo {pages} {3889} (\bibinfo {year} {2018})}\BibitemShut {NoStop}%
\bibitem [{\citenamefont {Winkel}\ \emph {et~al.}(2020)\citenamefont {Winkel}, \citenamefont {Borisov}, \citenamefont {Grünhaupt}, \citenamefont {Rieger}, \citenamefont {Spiecker}, \citenamefont {Valenti}, \citenamefont {Ustinov}, \citenamefont {Wernsdorfer},\ and\ \citenamefont {Pop}}]{winkelKerrGral2020}%
  \BibitemOpen
  \bibfield  {author} {\bibinfo {author} {\bibfnamefont {P.}~\bibnamefont {Winkel}}, \bibinfo {author} {\bibfnamefont {K.}~\bibnamefont {Borisov}}, \bibinfo {author} {\bibfnamefont {L.}~\bibnamefont {Grünhaupt}}, \bibinfo {author} {\bibfnamefont {D.}~\bibnamefont {Rieger}}, \bibinfo {author} {\bibfnamefont {M.}~\bibnamefont {Spiecker}}, \bibinfo {author} {\bibfnamefont {F.}~\bibnamefont {Valenti}}, \bibinfo {author} {\bibfnamefont {A.~V.}\ \bibnamefont {Ustinov}}, \bibinfo {author} {\bibfnamefont {W.}~\bibnamefont {Wernsdorfer}},\ and\ \bibinfo {author} {\bibfnamefont {I.~M.}\ \bibnamefont {Pop}},\ }\bibfield  {title} {\bibinfo {title} {Implementation of a transmon qubit using superconducting granular aluminum},\ }\href {https://doi.org/10.1103/PhysRevX.10.031032} {\bibfield  {journal} {\bibinfo  {journal} {Phys. Rev. X}\ }\textbf {\bibinfo {volume} {10}},\ \bibinfo {pages} {031032} (\bibinfo {year} {2020})}\BibitemShut {NoStop}%
\bibitem [{\citenamefont {Roy}\ \emph {et~al.}(2025)\citenamefont {Roy}, \citenamefont {Frasca},\ and\ \citenamefont {Scarlino}}]{Roy2025Mar}%
  \BibitemOpen
  \bibfield  {author} {\bibinfo {author} {\bibfnamefont {C.}~\bibnamefont {Roy}}, \bibinfo {author} {\bibfnamefont {S.}~\bibnamefont {Frasca}},\ and\ \bibinfo {author} {\bibfnamefont {P.}~\bibnamefont {Scarlino}},\ }\bibfield  {title} {\bibinfo {title} {{Study of Magnetic Field Resilient High Impedance High-Kinetic Inductance Superconducting Resonators}},\ }\bibfield  {journal} {\bibinfo  {journal} {arXiv}\ }\href {https://doi.org/10.48550/arXiv.2503.13321} {10.48550/arXiv.2503.13321} (\bibinfo {year} {2025}),\ \Eprint {https://arxiv.org/abs/2503.13321} {2503.13321} \BibitemShut {NoStop}%
\bibitem [{\citenamefont {Levy-Bertrand}\ \emph {et~al.}(2019)\citenamefont {Levy-Bertrand}, \citenamefont {Klein}, \citenamefont {Grenet}, \citenamefont {Dupré}, \citenamefont {Benoît}, \citenamefont {Bideaud}, \citenamefont {Bourrion}, \citenamefont {Calvo}, \citenamefont {Catalano}, \citenamefont {Gomez}, \citenamefont {Goupy}, \citenamefont {Grünhaupt}, \citenamefont {Luepke}, \citenamefont {Maleeva}, \citenamefont {Valenti}, \citenamefont {Pop},\ and\ \citenamefont {Monfardini}}]{levy-bertrandElectrodynamicgral2019}%
  \BibitemOpen
  \bibfield  {author} {\bibinfo {author} {\bibfnamefont {F.}~\bibnamefont {Levy-Bertrand}}, \bibinfo {author} {\bibfnamefont {T.}~\bibnamefont {Klein}}, \bibinfo {author} {\bibfnamefont {T.}~\bibnamefont {Grenet}}, \bibinfo {author} {\bibfnamefont {O.}~\bibnamefont {Dupré}}, \bibinfo {author} {\bibfnamefont {A.}~\bibnamefont {Benoît}}, \bibinfo {author} {\bibfnamefont {A.}~\bibnamefont {Bideaud}}, \bibinfo {author} {\bibfnamefont {O.}~\bibnamefont {Bourrion}}, \bibinfo {author} {\bibfnamefont {M.}~\bibnamefont {Calvo}}, \bibinfo {author} {\bibfnamefont {A.}~\bibnamefont {Catalano}}, \bibinfo {author} {\bibfnamefont {A.}~\bibnamefont {Gomez}}, \bibinfo {author} {\bibfnamefont {J.}~\bibnamefont {Goupy}}, \bibinfo {author} {\bibfnamefont {L.}~\bibnamefont {Grünhaupt}}, \bibinfo {author} {\bibfnamefont {U.~V.}\ \bibnamefont {Luepke}}, \bibinfo {author} {\bibfnamefont {N.}~\bibnamefont {Maleeva}}, \bibinfo {author} {\bibfnamefont {F.}~\bibnamefont {Valenti}}, \bibinfo {author} {\bibfnamefont {I.~M.}\
  \bibnamefont {Pop}},\ and\ \bibinfo {author} {\bibfnamefont {A.}~\bibnamefont {Monfardini}},\ }\bibfield  {title} {\bibinfo {title} {Electrodynamics of granular aluminum from superconductor to insulator: Observation of collective superconducting modes},\ }\href {https://doi.org/10.1103/PhysRevB.99.094506} {\bibfield  {journal} {\bibinfo  {journal} {Phys. Rev. B}\ }\textbf {\bibinfo {volume} {99}},\ \bibinfo {pages} {094506} (\bibinfo {year} {2019})}\BibitemShut {NoStop}%
\bibitem [{\citenamefont {Tinkham}(2004)}]{tinkhamintroduction2004}%
  \BibitemOpen
  \bibfield  {author} {\bibinfo {author} {\bibfnamefont {M.}~\bibnamefont {Tinkham}},\ }\href@noop {} {\emph {\bibinfo {title} {Introduction to Superconductivity}}}\ (\bibinfo {year} {2004})\BibitemShut {NoStop}%
\bibitem [{\citenamefont {Annunziata}\ \emph {et~al.}(2010)\citenamefont {Annunziata}, \citenamefont {Santavicca}, \citenamefont {Frunzio}, \citenamefont {Catelani}, \citenamefont {Rooks}, \citenamefont {Frydman},\ and\ \citenamefont {Prober}}]{annunziataSuperconducting2010}%
  \BibitemOpen
  \bibfield  {author} {\bibinfo {author} {\bibfnamefont {A.~J.}\ \bibnamefont {Annunziata}}, \bibinfo {author} {\bibfnamefont {D.~F.}\ \bibnamefont {Santavicca}}, \bibinfo {author} {\bibfnamefont {L.}~\bibnamefont {Frunzio}}, \bibinfo {author} {\bibfnamefont {G.}~\bibnamefont {Catelani}}, \bibinfo {author} {\bibfnamefont {M.~J.}\ \bibnamefont {Rooks}}, \bibinfo {author} {\bibfnamefont {A.}~\bibnamefont {Frydman}},\ and\ \bibinfo {author} {\bibfnamefont {D.~E.}\ \bibnamefont {Prober}},\ }\bibfield  {title} {\bibinfo {title} {Tunable superconducting nanoinductors},\ }\href {https://doi.org/10.1088/0957-4484/21/44/445202} {\bibfield  {journal} {\bibinfo  {journal} {Nanotechnology}\ }\textbf {\bibinfo {volume} {21}},\ \bibinfo {pages} {445202} (\bibinfo {year} {2010})}\BibitemShut {NoStop}%
\bibitem [{\citenamefont {Minev}\ \emph {et~al.}(2013)\citenamefont {Minev}, \citenamefont {Pop},\ and\ \citenamefont {Devoret}}]{minevPlanar2013}%
  \BibitemOpen
  \bibfield  {author} {\bibinfo {author} {\bibfnamefont {Z.~K.}\ \bibnamefont {Minev}}, \bibinfo {author} {\bibfnamefont {I.~M.}\ \bibnamefont {Pop}},\ and\ \bibinfo {author} {\bibfnamefont {M.~H.}\ \bibnamefont {Devoret}},\ }\bibfield  {title} {\bibinfo {title} {Planar superconducting whispering gallery mode resonators},\ }\href {https://doi.org/10.1063/1.4824201} {\bibfield  {journal} {\bibinfo  {journal} {Appl. Phys. Lett.}\ }\textbf {\bibinfo {volume} {103}},\ \bibinfo {pages} {142604} (\bibinfo {year} {2013})}\BibitemShut {NoStop}%
\bibitem [{\citenamefont {Pracht}\ \emph {et~al.}(2016)\citenamefont {Pracht}, \citenamefont {Bachar}, \citenamefont {Benfatto}, \citenamefont {Deutscher}, \citenamefont {Farber}, \citenamefont {Dressel},\ and\ \citenamefont {Scheffler}}]{prachtCooperPairing2016}%
  \BibitemOpen
  \bibfield  {author} {\bibinfo {author} {\bibfnamefont {U.~S.}\ \bibnamefont {Pracht}}, \bibinfo {author} {\bibfnamefont {N.}~\bibnamefont {Bachar}}, \bibinfo {author} {\bibfnamefont {L.}~\bibnamefont {Benfatto}}, \bibinfo {author} {\bibfnamefont {G.}~\bibnamefont {Deutscher}}, \bibinfo {author} {\bibfnamefont {E.}~\bibnamefont {Farber}}, \bibinfo {author} {\bibfnamefont {M.}~\bibnamefont {Dressel}},\ and\ \bibinfo {author} {\bibfnamefont {M.}~\bibnamefont {Scheffler}},\ }\bibfield  {title} {\bibinfo {title} {Enhanced {{Cooper}} pairing versus suppressed phase coherence shaping the superconducting dome in coupled aluminum nanograins},\ }\href {https://doi.org/10.1103/PhysRevB.93.100503} {\bibfield  {journal} {\bibinfo  {journal} {Phys. Rev. B}\ }\textbf {\bibinfo {volume} {93}},\ \bibinfo {pages} {100503} (\bibinfo {year} {2016})}\BibitemShut {NoStop}%
\bibitem [{\citenamefont {Henriques}\ \emph {et~al.}(2019)\citenamefont {Henriques}, \citenamefont {Valenti}, \citenamefont {Charpentier}, \citenamefont {Lagoin}, \citenamefont {Gouriou}, \citenamefont {Martínez}, \citenamefont {Cardani}, \citenamefont {Vignati}, \citenamefont {Grünhaupt}, \citenamefont {Gusenkova}, \citenamefont {Ferrero}, \citenamefont {Skacel}, \citenamefont {Wernsdorfer}, \citenamefont {Ustinov}, \citenamefont {Catelani}, \citenamefont {Sander},\ and\ \citenamefont {Pop}}]{henriquesPhonon2019}%
  \BibitemOpen
  \bibfield  {author} {\bibinfo {author} {\bibfnamefont {F.}~\bibnamefont {Henriques}}, \bibinfo {author} {\bibfnamefont {F.}~\bibnamefont {Valenti}}, \bibinfo {author} {\bibfnamefont {T.}~\bibnamefont {Charpentier}}, \bibinfo {author} {\bibfnamefont {M.}~\bibnamefont {Lagoin}}, \bibinfo {author} {\bibfnamefont {C.}~\bibnamefont {Gouriou}}, \bibinfo {author} {\bibfnamefont {M.}~\bibnamefont {Martínez}}, \bibinfo {author} {\bibfnamefont {L.}~\bibnamefont {Cardani}}, \bibinfo {author} {\bibfnamefont {M.}~\bibnamefont {Vignati}}, \bibinfo {author} {\bibfnamefont {L.}~\bibnamefont {Grünhaupt}}, \bibinfo {author} {\bibfnamefont {D.}~\bibnamefont {Gusenkova}}, \bibinfo {author} {\bibfnamefont {J.}~\bibnamefont {Ferrero}}, \bibinfo {author} {\bibfnamefont {S.~T.}\ \bibnamefont {Skacel}}, \bibinfo {author} {\bibfnamefont {W.}~\bibnamefont {Wernsdorfer}}, \bibinfo {author} {\bibfnamefont {A.~V.}\ \bibnamefont {Ustinov}}, \bibinfo {author} {\bibfnamefont {G.}~\bibnamefont {Catelani}}, \bibinfo {author} {\bibfnamefont
  {O.}~\bibnamefont {Sander}},\ and\ \bibinfo {author} {\bibfnamefont {I.~M.}\ \bibnamefont {Pop}},\ }\bibfield  {title} {\bibinfo {title} {Phonon traps reduce the quasiparticle density in superconducting circuits},\ }\href {https://doi.org/10.1063/1.5124967} {\bibfield  {journal} {\bibinfo  {journal} {Appl. Phys. Lett.}\ }\textbf {\bibinfo {volume} {115}},\ \bibinfo {pages} {212601} (\bibinfo {year} {2019})}\BibitemShut {NoStop}%
\bibitem [{\citenamefont {Grünhaupt}\ \emph {et~al.}(2017)\citenamefont {Grünhaupt}, \citenamefont {Von~Lüpke}, \citenamefont {Gusenkova}, \citenamefont {Skacel}, \citenamefont {Maleeva}, \citenamefont {Schlör}, \citenamefont {Bilmes}, \citenamefont {Rotzinger}, \citenamefont {Ustinov}, \citenamefont {Weides},\ and\ \citenamefont {Pop}}]{grunhauptArgonIonBeam2017}%
  \BibitemOpen
  \bibfield  {author} {\bibinfo {author} {\bibfnamefont {L.}~\bibnamefont {Grünhaupt}}, \bibinfo {author} {\bibfnamefont {U.}~\bibnamefont {Von~Lüpke}}, \bibinfo {author} {\bibfnamefont {D.}~\bibnamefont {Gusenkova}}, \bibinfo {author} {\bibfnamefont {S.~T.}\ \bibnamefont {Skacel}}, \bibinfo {author} {\bibfnamefont {N.}~\bibnamefont {Maleeva}}, \bibinfo {author} {\bibfnamefont {S.}~\bibnamefont {Schlör}}, \bibinfo {author} {\bibfnamefont {A.}~\bibnamefont {Bilmes}}, \bibinfo {author} {\bibfnamefont {H.}~\bibnamefont {Rotzinger}}, \bibinfo {author} {\bibfnamefont {A.~V.}\ \bibnamefont {Ustinov}}, \bibinfo {author} {\bibfnamefont {M.}~\bibnamefont {Weides}},\ and\ \bibinfo {author} {\bibfnamefont {I.~M.}\ \bibnamefont {Pop}},\ }\bibfield  {title} {\bibinfo {title} {An argon ion beam milling process for native alox layers enabling coherent superconducting contacts},\ }\href {https://doi.org/10.1063/1.4990491} {\bibfield  {journal} {\bibinfo  {journal} {Phys. Rev. Appl.}\ }\textbf {\bibinfo {volume} {111}},\
  \bibinfo {pages} {072601} (\bibinfo {year} {2017})}\BibitemShut {NoStop}%
\bibitem [{\citenamefont {Bartlett}(1948)}]{bartlettnsd1948}%
  \BibitemOpen
  \bibfield  {author} {\bibinfo {author} {\bibfnamefont {M.~S.}\ \bibnamefont {Bartlett}},\ }\bibfield  {title} {\bibinfo {title} {Smoothing periodograms from time-series with continuous spectra},\ }\href {https://doi.org/10.1038/161686a0} {\bibfield  {journal} {\bibinfo  {journal} {Nature}\ }\textbf {\bibinfo {volume} {161}},\ \bibinfo {pages} {686} (\bibinfo {year} {1948})}\BibitemShut {NoStop}%
\bibitem [{\citenamefont {Valenti}\ \emph {et~al.}(2019)\citenamefont {Valenti}, \citenamefont {Henriques}, \citenamefont {Catelani}, \citenamefont {Maleeva}, \citenamefont {Grünhaupt}, \citenamefont {Von~Lüpke}, \citenamefont {Skacel}, \citenamefont {Winkel}, \citenamefont {Bilmes}, \citenamefont {Ustinov}, \citenamefont {Goupy}, \citenamefont {Calvo}, \citenamefont {Benoît}, \citenamefont {Levy-Bertrand}, \citenamefont {Monfardini},\ and\ \citenamefont {Pop}}]{valentiKineticInductance2019}%
  \BibitemOpen
  \bibfield  {author} {\bibinfo {author} {\bibfnamefont {F.}~\bibnamefont {Valenti}}, \bibinfo {author} {\bibfnamefont {F.}~\bibnamefont {Henriques}}, \bibinfo {author} {\bibfnamefont {G.}~\bibnamefont {Catelani}}, \bibinfo {author} {\bibfnamefont {N.}~\bibnamefont {Maleeva}}, \bibinfo {author} {\bibfnamefont {L.}~\bibnamefont {Grünhaupt}}, \bibinfo {author} {\bibfnamefont {U.}~\bibnamefont {Von~Lüpke}}, \bibinfo {author} {\bibfnamefont {S.~T.}\ \bibnamefont {Skacel}}, \bibinfo {author} {\bibfnamefont {P.}~\bibnamefont {Winkel}}, \bibinfo {author} {\bibfnamefont {A.}~\bibnamefont {Bilmes}}, \bibinfo {author} {\bibfnamefont {A.~V.}\ \bibnamefont {Ustinov}}, \bibinfo {author} {\bibfnamefont {J.}~\bibnamefont {Goupy}}, \bibinfo {author} {\bibfnamefont {M.}~\bibnamefont {Calvo}}, \bibinfo {author} {\bibfnamefont {A.}~\bibnamefont {Benoît}}, \bibinfo {author} {\bibfnamefont {F.}~\bibnamefont {Levy-Bertrand}}, \bibinfo {author} {\bibfnamefont {A.}~\bibnamefont {Monfardini}},\ and\ \bibinfo {author}
  {\bibfnamefont {I.~M.}\ \bibnamefont {Pop}},\ }\bibfield  {title} {\bibinfo {title} {Interplay between kinetic inductance, nonlinearity, and quasiparticle dynamics in granular aluminum microwave kinetic inductance detectors},\ }\href {https://doi.org/10.1103/PhysRevApplied.11.054087} {\bibfield  {journal} {\bibinfo  {journal} {Phys. Rev. Appl.}\ }\textbf {\bibinfo {volume} {11}},\ \bibinfo {pages} {054087} (\bibinfo {year} {2019})}\BibitemShut {NoStop}%
\bibitem [{\citenamefont {Burnett}\ \emph {et~al.}(2018)\citenamefont {Burnett}, \citenamefont {Bengtsson}, \citenamefont {Niepce},\ and\ \citenamefont {Bylander}}]{burnettNoise2018}%
  \BibitemOpen
  \bibfield  {author} {\bibinfo {author} {\bibfnamefont {J.}~\bibnamefont {Burnett}}, \bibinfo {author} {\bibfnamefont {A.}~\bibnamefont {Bengtsson}}, \bibinfo {author} {\bibfnamefont {D.}~\bibnamefont {Niepce}},\ and\ \bibinfo {author} {\bibfnamefont {J.}~\bibnamefont {Bylander}},\ }\bibfield  {title} {\bibinfo {title} {Noise and loss of superconducting aluminium resonators at single photon energies},\ }\href {https://doi.org/10.1088/1742-6596/969/1/012131} {\bibfield  {journal} {\bibinfo  {journal} {J. Phys. Conf. Ser.}\ }\textbf {\bibinfo {volume} {969}},\ \bibinfo {pages} {012131} (\bibinfo {year} {2018})}\BibitemShut {NoStop}%
\bibitem [{\citenamefont {Schlör}\ \emph {et~al.}(2019)\citenamefont {Schlör}, \citenamefont {Lisenfeld}, \citenamefont {Müller}, \citenamefont {Bilmes}, \citenamefont {Schneider}, \citenamefont {Pappas}, \citenamefont {Ustinov},\ and\ \citenamefont {Weides}}]{schlornoise2019}%
  \BibitemOpen
  \bibfield  {author} {\bibinfo {author} {\bibfnamefont {S.}~\bibnamefont {Schlör}}, \bibinfo {author} {\bibfnamefont {J.}~\bibnamefont {Lisenfeld}}, \bibinfo {author} {\bibfnamefont {C.}~\bibnamefont {Müller}}, \bibinfo {author} {\bibfnamefont {A.}~\bibnamefont {Bilmes}}, \bibinfo {author} {\bibfnamefont {A.}~\bibnamefont {Schneider}}, \bibinfo {author} {\bibfnamefont {D.~P.}\ \bibnamefont {Pappas}}, \bibinfo {author} {\bibfnamefont {A.~V.}\ \bibnamefont {Ustinov}},\ and\ \bibinfo {author} {\bibfnamefont {M.}~\bibnamefont {Weides}},\ }\bibfield  {title} {\bibinfo {title} {Correlating decoherence in transmon qubits: Low frequency noise by single fluctuators},\ }\href {https://doi.org/10.1103/PhysRevLett.123.190502} {\bibfield  {journal} {\bibinfo  {journal} {Phys. Rev. Lett.}\ }\textbf {\bibinfo {volume} {123}},\ \bibinfo {pages} {190502} (\bibinfo {year} {2019})}\BibitemShut {NoStop}%
\bibitem [{\citenamefont {Kristen}\ \emph {et~al.}(2023)\citenamefont {Kristen}, \citenamefont {Voss}, \citenamefont {Wildermuth}, \citenamefont {Rotzinger},\ and\ \citenamefont {Ustinov}}]{kristenFluctuations2023}%
  \BibitemOpen
  \bibfield  {author} {\bibinfo {author} {\bibfnamefont {M.}~\bibnamefont {Kristen}}, \bibinfo {author} {\bibfnamefont {J.~N.}\ \bibnamefont {Voss}}, \bibinfo {author} {\bibfnamefont {M.}~\bibnamefont {Wildermuth}}, \bibinfo {author} {\bibfnamefont {H.}~\bibnamefont {Rotzinger}},\ and\ \bibinfo {author} {\bibfnamefont {A.~V.}\ \bibnamefont {Ustinov}},\ }\bibfield  {title} {\bibinfo {title} {Random telegraph fluctuations in granular microwave resonators},\ }\href {https://doi.org/10.1063/5.0147430} {\bibfield  {journal} {\bibinfo  {journal} {Phys. Rev. Appl.}\ }\textbf {\bibinfo {volume} {122}},\ \bibinfo {pages} {202602} (\bibinfo {year} {2023})}\BibitemShut {NoStop}%
\bibitem [{\citenamefont {Rieger}\ \emph {et~al.}(2023)\citenamefont {Rieger}, \citenamefont {Günzler}, \citenamefont {Spiecker}, \citenamefont {Nambisan}, \citenamefont {Wernsdorfer},\ and\ \citenamefont {Pop}}]{riegerFano2023}%
  \BibitemOpen
  \bibfield  {author} {\bibinfo {author} {\bibfnamefont {D.}~\bibnamefont {Rieger}}, \bibinfo {author} {\bibfnamefont {S.}~\bibnamefont {Günzler}}, \bibinfo {author} {\bibfnamefont {M.}~\bibnamefont {Spiecker}}, \bibinfo {author} {\bibfnamefont {A.}~\bibnamefont {Nambisan}}, \bibinfo {author} {\bibfnamefont {W.}~\bibnamefont {Wernsdorfer}},\ and\ \bibinfo {author} {\bibfnamefont {I.~M.}\ \bibnamefont {Pop}},\ }\bibfield  {title} {\bibinfo {title} {Fano interference in microwave resonator measurements},\ }\href {https://doi.org/10.1103/PhysRevApplied.20.014059} {\bibfield  {journal} {\bibinfo  {journal} {Phys. Rev. Appl.}\ }\textbf {\bibinfo {volume} {20}},\ \bibinfo {pages} {014059} (\bibinfo {year} {2023})}\BibitemShut {NoStop}%
\bibitem [{\citenamefont {Grünhaupt}\ \emph {et~al.}(2018)\citenamefont {Grünhaupt}, \citenamefont {Maleeva}, \citenamefont {Skacel}, \citenamefont {Calvo}, \citenamefont {Levy-Bertrand}, \citenamefont {Ustinov}, \citenamefont {Rotzinger}, \citenamefont {Monfardini}, \citenamefont {Catelani},\ and\ \citenamefont {Pop}}]{grunhauptgralLoss2018}%
  \BibitemOpen
  \bibfield  {author} {\bibinfo {author} {\bibfnamefont {L.}~\bibnamefont {Grünhaupt}}, \bibinfo {author} {\bibfnamefont {N.}~\bibnamefont {Maleeva}}, \bibinfo {author} {\bibfnamefont {S.~T.}\ \bibnamefont {Skacel}}, \bibinfo {author} {\bibfnamefont {M.}~\bibnamefont {Calvo}}, \bibinfo {author} {\bibfnamefont {F.}~\bibnamefont {Levy-Bertrand}}, \bibinfo {author} {\bibfnamefont {A.~V.}\ \bibnamefont {Ustinov}}, \bibinfo {author} {\bibfnamefont {H.}~\bibnamefont {Rotzinger}}, \bibinfo {author} {\bibfnamefont {A.}~\bibnamefont {Monfardini}}, \bibinfo {author} {\bibfnamefont {G.}~\bibnamefont {Catelani}},\ and\ \bibinfo {author} {\bibfnamefont {I.~M.}\ \bibnamefont {Pop}},\ }\bibfield  {title} {\bibinfo {title} {Loss mechanisms and quasiparticle dynamics in superconducting microwave resonators made of thin-film granular aluminum},\ }\href {https://doi.org/10.1103/PhysRevLett.121.117001} {\bibfield  {journal} {\bibinfo  {journal} {Phys. Rev. Lett.}\ }\textbf {\bibinfo {volume} {121}},\ \bibinfo {pages} {117001}
  (\bibinfo {year} {2018})}\BibitemShut {NoStop}%
\bibitem [{\citenamefont {Hunklinger}\ \emph {et~al.}(1972)\citenamefont {Hunklinger}, \citenamefont {Arnold}, \citenamefont {Stein}, \citenamefont {Nava},\ and\ \citenamefont {Dransfeld}}]{hunklingerSaturation1972}%
  \BibitemOpen
  \bibfield  {author} {\bibinfo {author} {\bibfnamefont {S.}~\bibnamefont {Hunklinger}}, \bibinfo {author} {\bibfnamefont {W.}~\bibnamefont {Arnold}}, \bibinfo {author} {\bibfnamefont {S.}~\bibnamefont {Stein}}, \bibinfo {author} {\bibfnamefont {R.}~\bibnamefont {Nava}},\ and\ \bibinfo {author} {\bibfnamefont {K.}~\bibnamefont {Dransfeld}},\ }\bibfield  {title} {\bibinfo {title} {Saturation of the ultrasonic absorption in vitreous silica at low temperatures},\ }\href {https://doi.org/10.1016/0375-9601(72)90884-5} {\bibfield  {journal} {\bibinfo  {journal} {Phys. Lett. A}\ }\textbf {\bibinfo {volume} {42}},\ \bibinfo {pages} {253} (\bibinfo {year} {1972})}\BibitemShut {NoStop}%
\bibitem [{\citenamefont {Golding}\ \emph {et~al.}(1973)\citenamefont {Golding}, \citenamefont {Graebner}, \citenamefont {Halperin},\ and\ \citenamefont {Schutz}}]{goldingNonlinearPhonon1973}%
  \BibitemOpen
  \bibfield  {author} {\bibinfo {author} {\bibfnamefont {B.}~\bibnamefont {Golding}}, \bibinfo {author} {\bibfnamefont {J.~E.}\ \bibnamefont {Graebner}}, \bibinfo {author} {\bibfnamefont {B.~I.}\ \bibnamefont {Halperin}},\ and\ \bibinfo {author} {\bibfnamefont {R.~J.}\ \bibnamefont {Schutz}},\ }\bibfield  {title} {\bibinfo {title} {Nonlinear phonon propagation in fused silica below 1 k},\ }\href {https://doi.org/10.1103/PhysRevLett.30.223} {\bibfield  {journal} {\bibinfo  {journal} {Phys. Rev. Lett.}\ }\textbf {\bibinfo {volume} {30}},\ \bibinfo {pages} {223} (\bibinfo {year} {1973})}\BibitemShut {NoStop}%
\bibitem [{\citenamefont {Levenson-Falk}\ \emph {et~al.}(2014)\citenamefont {Levenson-Falk}, \citenamefont {Kos}, \citenamefont {Vijay}, \citenamefont {Glazman},\ and\ \citenamefont {Siddiqi}}]{levensoneQuasiparticle2014}%
  \BibitemOpen
  \bibfield  {author} {\bibinfo {author} {\bibfnamefont {E.~M.}\ \bibnamefont {Levenson-Falk}}, \bibinfo {author} {\bibfnamefont {F.}~\bibnamefont {Kos}}, \bibinfo {author} {\bibfnamefont {R.}~\bibnamefont {Vijay}}, \bibinfo {author} {\bibfnamefont {L.}~\bibnamefont {Glazman}},\ and\ \bibinfo {author} {\bibfnamefont {I.}~\bibnamefont {Siddiqi}},\ }\bibfield  {title} {\bibinfo {title} {Single-quasiparticle trapping in aluminum nanobridge josephson junctions},\ }\href {https://doi.org/10.1103/PhysRevLett.112.047002} {\bibfield  {journal} {\bibinfo  {journal} {Phys. Rev. Lett.}\ }\textbf {\bibinfo {volume} {112}},\ \bibinfo {pages} {047002} (\bibinfo {year} {2014})}\BibitemShut {NoStop}%
\bibitem [{\citenamefont {Gustavsson}\ \emph {et~al.}(2016)\citenamefont {Gustavsson}, \citenamefont {Yan}, \citenamefont {Catelani}, \citenamefont {Bylander}, \citenamefont {Kamal}, \citenamefont {Birenbaum}, \citenamefont {Hover}, \citenamefont {Rosenberg}, \citenamefont {Samach}, \citenamefont {Sears}, \citenamefont {Weber}, \citenamefont {Yoder}, \citenamefont {Clarke}, \citenamefont {Kerman}, \citenamefont {Yoshihara}, \citenamefont {Nakamura}, \citenamefont {Orlando},\ and\ \citenamefont {Oliver}}]{gustavssonRelaxation2016}%
  \BibitemOpen
  \bibfield  {author} {\bibinfo {author} {\bibfnamefont {S.}~\bibnamefont {Gustavsson}}, \bibinfo {author} {\bibfnamefont {F.}~\bibnamefont {Yan}}, \bibinfo {author} {\bibfnamefont {G.}~\bibnamefont {Catelani}}, \bibinfo {author} {\bibfnamefont {J.}~\bibnamefont {Bylander}}, \bibinfo {author} {\bibfnamefont {A.}~\bibnamefont {Kamal}}, \bibinfo {author} {\bibfnamefont {J.}~\bibnamefont {Birenbaum}}, \bibinfo {author} {\bibfnamefont {D.}~\bibnamefont {Hover}}, \bibinfo {author} {\bibfnamefont {D.}~\bibnamefont {Rosenberg}}, \bibinfo {author} {\bibfnamefont {G.}~\bibnamefont {Samach}}, \bibinfo {author} {\bibfnamefont {A.~P.}\ \bibnamefont {Sears}}, \bibinfo {author} {\bibfnamefont {S.~J.}\ \bibnamefont {Weber}}, \bibinfo {author} {\bibfnamefont {J.~L.}\ \bibnamefont {Yoder}}, \bibinfo {author} {\bibfnamefont {J.}~\bibnamefont {Clarke}}, \bibinfo {author} {\bibfnamefont {A.~J.}\ \bibnamefont {Kerman}}, \bibinfo {author} {\bibfnamefont {F.}~\bibnamefont {Yoshihara}}, \bibinfo {author} {\bibfnamefont
  {Y.}~\bibnamefont {Nakamura}}, \bibinfo {author} {\bibfnamefont {T.~P.}\ \bibnamefont {Orlando}},\ and\ \bibinfo {author} {\bibfnamefont {W.~D.}\ \bibnamefont {Oliver}},\ }\bibfield  {title} {\bibinfo {title} {Suppressing relaxation in superconducting qubits by quasiparticle pumping},\ }\href {https://doi.org/10.1126/science.aah5844} {\bibfield  {journal} {\bibinfo  {journal} {Science}\ }\textbf {\bibinfo {volume} {354}},\ \bibinfo {pages} {1573} (\bibinfo {year} {2016})}\BibitemShut {NoStop}%
\bibitem [{\citenamefont {Jin}\ \emph {et~al.}(2025)\citenamefont {Jin}, \citenamefont {Serpico}, \citenamefont {Lee}, \citenamefont {Confalone}, \citenamefont {Saggau}, \citenamefont {Lo~Sardo}, \citenamefont {Gu}, \citenamefont {Goodge}, \citenamefont {Lesne}, \citenamefont {Montemurro}, \citenamefont {Nielsch}, \citenamefont {Poccia},\ and\ \citenamefont {Vool}}]{jinExplorings2025}%
  \BibitemOpen
  \bibfield  {author} {\bibinfo {author} {\bibfnamefont {H.}~\bibnamefont {Jin}}, \bibinfo {author} {\bibfnamefont {G.}~\bibnamefont {Serpico}}, \bibinfo {author} {\bibfnamefont {Y.}~\bibnamefont {Lee}}, \bibinfo {author} {\bibfnamefont {T.}~\bibnamefont {Confalone}}, \bibinfo {author} {\bibfnamefont {C.~N.}\ \bibnamefont {Saggau}}, \bibinfo {author} {\bibfnamefont {F.}~\bibnamefont {Lo~Sardo}}, \bibinfo {author} {\bibfnamefont {G.}~\bibnamefont {Gu}}, \bibinfo {author} {\bibfnamefont {B.~H.}\ \bibnamefont {Goodge}}, \bibinfo {author} {\bibfnamefont {E.}~\bibnamefont {Lesne}}, \bibinfo {author} {\bibfnamefont {D.}~\bibnamefont {Montemurro}}, \bibinfo {author} {\bibfnamefont {K.}~\bibnamefont {Nielsch}}, \bibinfo {author} {\bibfnamefont {N.}~\bibnamefont {Poccia}},\ and\ \bibinfo {author} {\bibfnamefont {U.}~\bibnamefont {Vool}},\ }\bibfield  {title} {\bibinfo {title} {Exploring van der {{Waals}} cuprate superconductors using a hybrid microwave circuit},\ }\href {https://doi.org/10.1021/acs.nanolett.4c05793}
  {\bibfield  {journal} {\bibinfo  {journal} {Nano Lett.}\ } (\bibinfo {year} {2025})}\BibitemShut {NoStop}%
\bibitem [{\citenamefont {Andersson}\ \emph {et~al.}(2021)\citenamefont {Andersson}, \citenamefont {Bilobran}, \citenamefont {Scigliuzzo}, \citenamefont {de~Lima}, \citenamefont {Cole},\ and\ \citenamefont {Delsing}}]{Andersson2021Jan}%
  \BibitemOpen
  \bibfield  {author} {\bibinfo {author} {\bibfnamefont {G.}~\bibnamefont {Andersson}}, \bibinfo {author} {\bibfnamefont {A.~L.~O.}\ \bibnamefont {Bilobran}}, \bibinfo {author} {\bibfnamefont {M.}~\bibnamefont {Scigliuzzo}}, \bibinfo {author} {\bibfnamefont {M.~M.}\ \bibnamefont {de~Lima}}, \bibinfo {author} {\bibfnamefont {J.~H.}\ \bibnamefont {Cole}},\ and\ \bibinfo {author} {\bibfnamefont {P.}~\bibnamefont {Delsing}},\ }\bibfield  {title} {\bibinfo {title} {Acoustic spectral hole-burning in a two-level system ensemble},\ }\href {https://doi.org/10.1038/s41534-020-00348-0} {\bibfield  {journal} {\bibinfo  {journal} {npj Quantum Inf.}\ }\textbf {\bibinfo {volume} {7}},\ \bibinfo {pages} {1} (\bibinfo {year} {2021})}\BibitemShut {NoStop}%
\bibitem [{\citenamefont {Capelle}\ \emph {et~al.}(2020)\citenamefont {Capelle}, \citenamefont {Flurin}, \citenamefont {Ivanov}, \citenamefont {Palomo}, \citenamefont {Rosticher}, \citenamefont {Chua}, \citenamefont {Briant}, \citenamefont {Cohadon}, \citenamefont {Heidmann}, \citenamefont {Jacqmin},\ and\ \citenamefont {Del{\ifmmode\acute{e}\else\'{e}\fi}glise}}]{Capelle2020Mar}%
  \BibitemOpen
  \bibfield  {author} {\bibinfo {author} {\bibfnamefont {T.}~\bibnamefont {Capelle}}, \bibinfo {author} {\bibfnamefont {E.}~\bibnamefont {Flurin}}, \bibinfo {author} {\bibfnamefont {E.}~\bibnamefont {Ivanov}}, \bibinfo {author} {\bibfnamefont {J.}~\bibnamefont {Palomo}}, \bibinfo {author} {\bibfnamefont {M.}~\bibnamefont {Rosticher}}, \bibinfo {author} {\bibfnamefont {S.}~\bibnamefont {Chua}}, \bibinfo {author} {\bibfnamefont {T.}~\bibnamefont {Briant}}, \bibinfo {author} {\bibfnamefont {P.-F.}\ \bibnamefont {Cohadon}}, \bibinfo {author} {\bibfnamefont {A.}~\bibnamefont {Heidmann}}, \bibinfo {author} {\bibfnamefont {T.}~\bibnamefont {Jacqmin}},\ and\ \bibinfo {author} {\bibfnamefont {S.}~\bibnamefont {Del{\ifmmode\acute{e}\else\'{e}\fi}glise}},\ }\bibfield  {title} {\bibinfo {title} {Probing a two-level system bath via the frequency shift of an off-resonantly driven cavity},\ }\href {https://doi.org/10.1103/PhysRevApplied.13.034022} {\bibfield  {journal} {\bibinfo  {journal} {Phys. Rev. Appl.}\ }\textbf
  {\bibinfo {volume} {13}},\ \bibinfo {pages} {034022} (\bibinfo {year} {2020})}\BibitemShut {NoStop}%
\bibitem [{\citenamefont {Kirsh}\ \emph {et~al.}(2017)\citenamefont {Kirsh}, \citenamefont {Svetitsky}, \citenamefont {Burin}, \citenamefont {Schechter},\ and\ \citenamefont {Katz}}]{Kirsh2017Jun}%
  \BibitemOpen
  \bibfield  {author} {\bibinfo {author} {\bibfnamefont {N.}~\bibnamefont {Kirsh}}, \bibinfo {author} {\bibfnamefont {E.}~\bibnamefont {Svetitsky}}, \bibinfo {author} {\bibfnamefont {A.~L.}\ \bibnamefont {Burin}}, \bibinfo {author} {\bibfnamefont {M.}~\bibnamefont {Schechter}},\ and\ \bibinfo {author} {\bibfnamefont {N.}~\bibnamefont {Katz}},\ }\bibfield  {title} {\bibinfo {title} {Revealing the nonlinear response of a tunneling two-level system ensemble using coupled modes},\ }\href {https://doi.org/10.1103/PhysRevMaterials.1.012601} {\bibfield  {journal} {\bibinfo  {journal} {Phys. Rev. Mater.}\ }\textbf {\bibinfo {volume} {1}},\ \bibinfo {pages} {012601} (\bibinfo {year} {2017})}\BibitemShut {NoStop}%
\bibitem [{\citenamefont {Zypman}(2019)}]{Zypman2019Sep}%
  \BibitemOpen
  \bibfield  {author} {\bibinfo {author} {\bibfnamefont {F.~R.}\ \bibnamefont {Zypman}},\ }\bibfield  {title} {\bibinfo {title} {Mathematical expression for the capacitance of coplanar strips},\ }\href {https://doi.org/10.1016/j.elstat.2019.103371} {\bibfield  {journal} {\bibinfo  {journal} {J. Electrostat.}\ }\textbf {\bibinfo {volume} {101}},\ \bibinfo {pages} {103371} (\bibinfo {year} {2019})}\BibitemShut {NoStop}%
\end{thebibliography}%
\balancecolsandclearpage

\end{document}